\documentclass[fleqn,usenatbib,useAMS]{mnras}  

\usepackage[T1]{fontenc}
\usepackage{ae,aecompl}
\usepackage{graphicx,amsmath,amssymb}

\newcommand \radm{rad~m$^{-2}$}

\title[Finding a complex polarized radio signal]{Finding a complex polarized signal in wide-band radio data}

\author[D.~H.~F.~M. Schnitzeler ]{
D.~H.~F.~M. Schnitzeler$^{1,2}$\thanks{dschnitzeler@gmail.com}
\\ 
$^1$ Max Planck Institut f\"ur Radioastronomie, 53121 Bonn, Germany\\
$^2$ Bendenweg 51, 53121 Bonn, Germany\\
}

\date{Accepted 2017 October 19 . Received 2017 October 19 ; in original form 2017 August 22}

\pubyear{}

\begin{document}
\label{firstpage}
\pagerange{\pageref{firstpage}--\pageref{lastpage}}
\maketitle

\begin{abstract}
We present a new algorithm for fitting and classifying polarized radio sources, which is based on the QU fitting method introduced by O'Sullivan et al. and on our analysis of pulsars.
Then we test this algorithm using Monte Carlo simulations of observations in the 16 cm band of the Australia Telescope Compact Array (1.3-3.1 GHz), to quantify how often the algorithm identifies the correct source model, how certain it is of this identification, and how the parameters of the injected and fitted models compare. 
In our analysis we consider the Akaike and Bayesian Information Criteria, and model averaging. For the observing setup we simulated, the Bayesian Information Criterion, without model averaging, is the best way for identifying the correct model and for estimating its parameters. Sources can only be identified correctly if their parameters lie inside a `Goldilocks region': strong depolarization makes it impossible to detect sources that emit over a wide range in RM, whereas sources that emit over a narrow range in RM cannot be told apart from simpler sources or sources that emit at only one RM. We identify when emission at similar RMs is `resolved', and quantify this in a way similar to the Rayleigh criterion in optics.
Also, we identify pitfalls in RM synthesis that are avoided by QU fitting.
Finally, we show how channel weights can be tweaked to produce apodized RM spectra, that observing time requirements in RM synthesis and QU fitting are the same, and we analyse when to stop RMClean.
\end{abstract}

%

\section{Introduction}\label{introduction.sec}
Understanding the role of magnetic fields in astrophysics, and how these fields originated, critically depends on our ability to extract useful information from data sets which are often noisy, the topic of our paper.
Faraday rotation of linearly polarized radio waves enables us to study magnetic fields:  
the Faraday effect rotates the plane of polarization of waves passing through an ionized medium with an embedded magnetic field.
If the source emits radio waves with their plane of polarization at an angle $\chi_0$, then the magnitude of the Faraday effect is described by $\chi-\chi_0=\mathrm{RM}\lambda^2$, where $\chi$ is the observed polarization angle of the radio waves, $\lambda$ the observing wavelength, and the rotation measure RM depends on the physical properties of the magnetized plasma:
\begin{eqnarray}
\mathrm{RM}\, \left(\mathrm{rad~m}^{-2}\right)\ \approx\ 0.81 \int_\mathrm{source}^\mathrm{observer} n_\mathrm{e} B_\| \mathrm{d}l\, .
\label{rm_definition.eqn}
\end{eqnarray}
Here $n_\mathrm{e}$ is the free electron density (cm$^{-3}$), $B_\|$ the length of the magnetic field vector projected along the line of sight ($\mu$G), and d$l$ an infinitesimal distance interval along the line of sight from the source to the observer (pc). We will follow the naming convention presented in \citet[`SL17']{schnitzeler2017}.
Faraday rotation can manifest itself not only as a change in the polarization angle with frequency, but also as a change in the polarized flux density (depolarization) with frequency (e.g., \citealt{kuzmin1959}, \citealt{woltjer1962}, \citealt{morris1963}, \citealt{gardner1963}). 

With the advent of broad-band receivers it has become possible to accurately measure the linear Stokes parameters $Q$ and $U$ across a wide range of frequencies in a short period of time. 
The two most popular techniques for extracting information from these frequency spectra are RM synthesis (\citealt{burn1966}, \citealt{brentjens2005}) and QU fitting (\citealt{farnsworth2011}, \citealt{osullivan2012}); additional methods have been discussed by \cite{sun2015}. 

To extract information from wide-band data sets, two effects need to be taken into account.
First, sources of synchrotron radiation have spectral indices that are typically not zero, which means that the polarized flux density of these sources changes with frequency even if there is no depolarization.
Second, the sensitivity of the measurements can change across the observing band due to changes in the system temperature with frequency or because data points had to be flagged non-uniformly across the frequency band.

\begin{figure}
\begin{centering}
\resizebox{0.9\hsize}{!}{\includegraphics{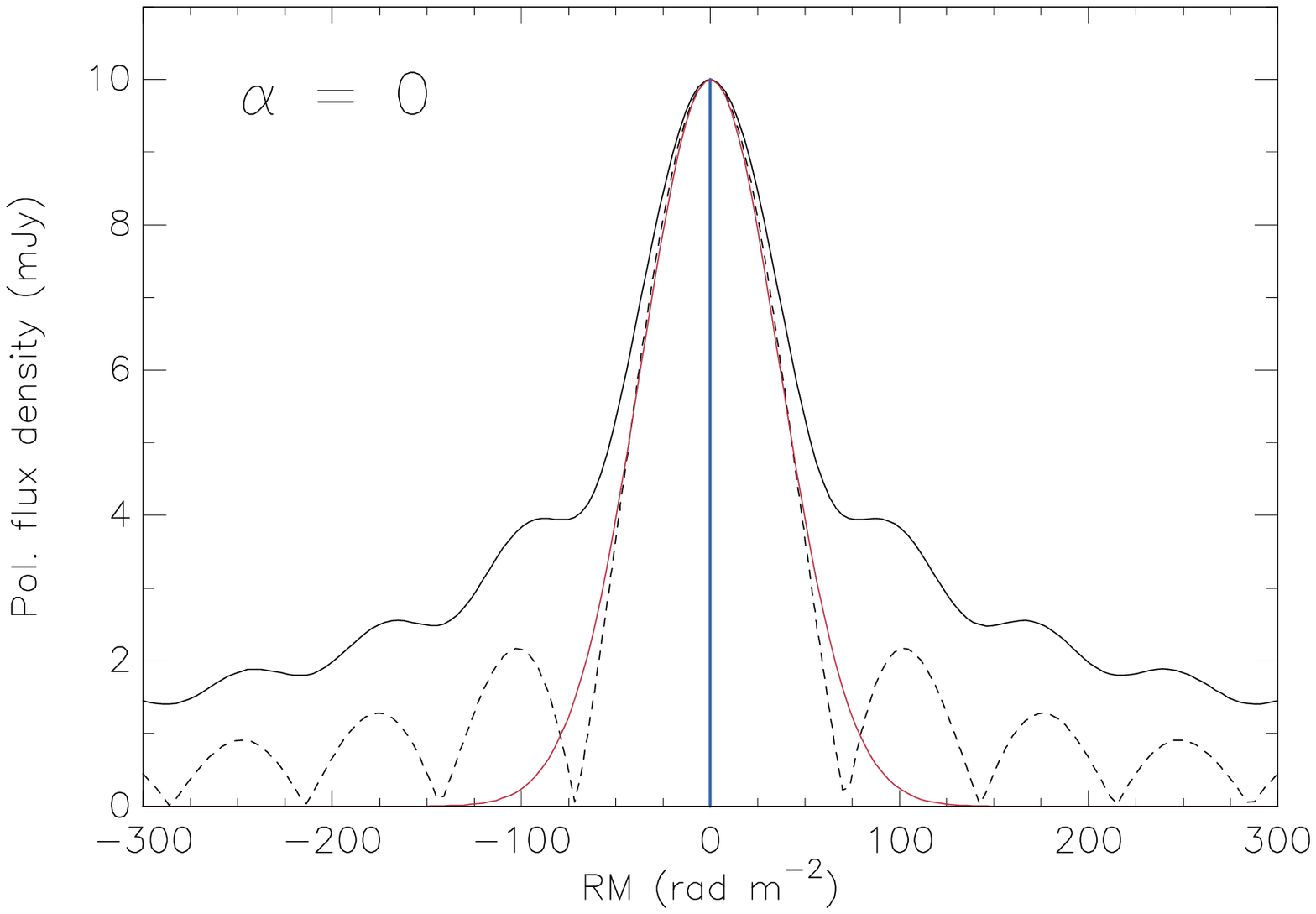}}
\resizebox{0.9\hsize}{!}{\includegraphics{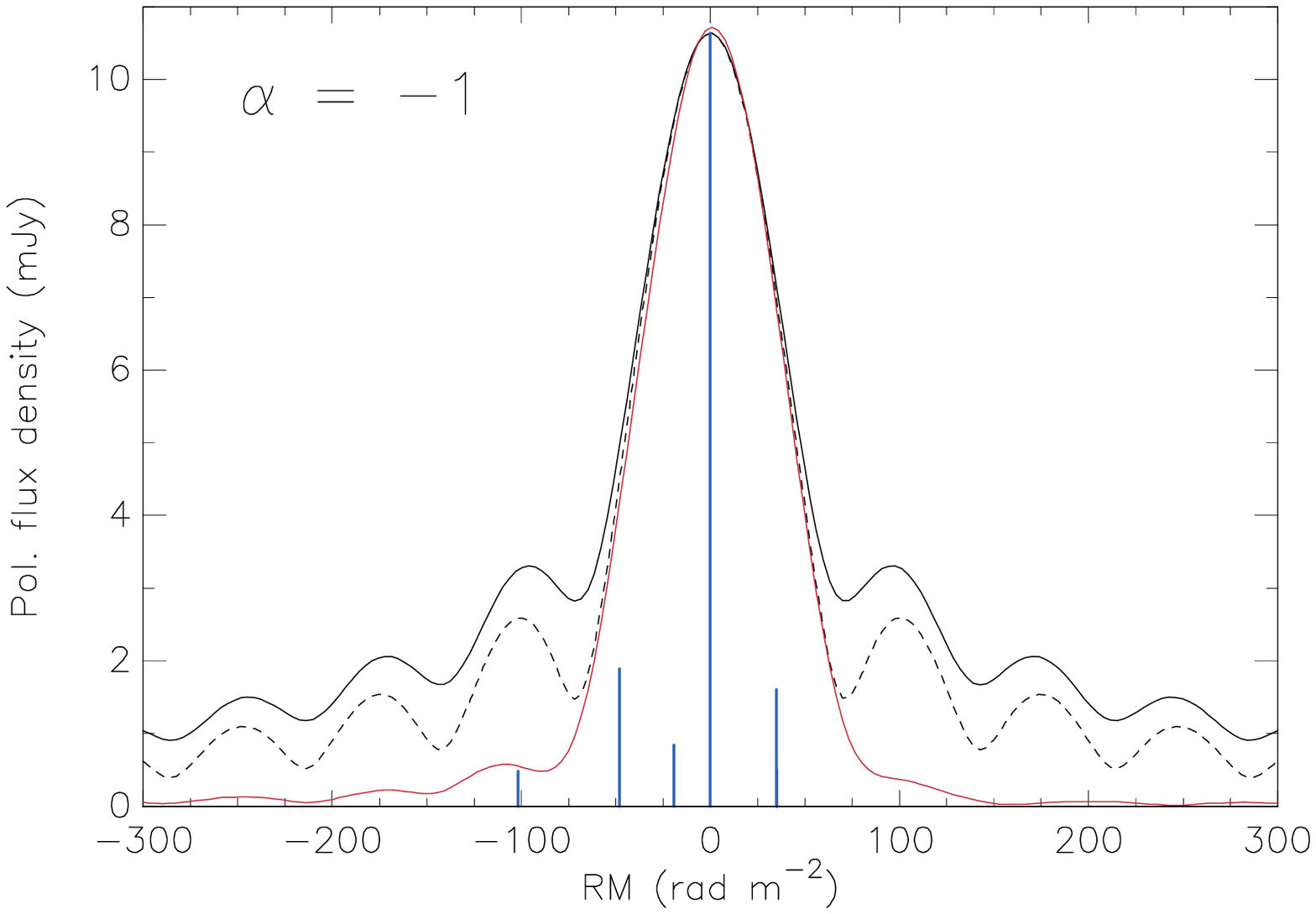}}
\caption{
Example of how non-zero spectral indices can produce artefacts in cleaned RM spectra. To create these figures, we simulated a source that emits 10 units of flux density at a frequency of 2.2 GHz; the polarized emission from this source has a spectral index $\alpha$ = 0 (top panel) or $\alpha$ = -1 (bottom panel) and a rotation measure $\mathrm{RM}=0$ \radm.
We simulated observations between 1.3-3.1 GHz, and applied RM synthesis and RMClean to these observations.
Both panels show raw and cleaned RM spectra (black and red lines, respectively) together with all clean components (blue vertical lines). 
The dashed line shows the raw RM spectrum for an observation in wavelength squared instead of frequency, using identical coverages in wavelength squared and identical numbers of channels in both cases.
}
\label{steepsp.fig}
\end{centering}
\end{figure}

To demonstrate that spectral index effects can have an important impact on data analysis, even leading to artefacts, we generated two mock data sets for observations between 1.3-3.1 GHz, and analysed these data sets using RM synthesis and RMClean \citep{heald2009}, which is the standard deconvolution algorithm for RM spectra.
At the beginning of Section~\ref{spass.sec} we describe the details of this frequency setup.
Each data set contains a source that emits a polarized flux density $L\left(\nu\right) = 10\,\mathrm{mJy}\left(\nu/2196\, \mathrm{MHz}\right)^\alpha$, with $\alpha = 0$ or -1, and has an RM of zero \radm\
(throughout this paper, we will use `$\alpha$' for the spectral index in polarized flux density).
The results from our analysis are shown in Fig.~\ref{steepsp.fig}.
The top panel of this figure shows polarized emission only at zero \radm , as one would expect for this source, but the bottom panel shows additional components that are all spurious.
When deconvolving RM spectra, RMClean has to assume a certain shape for the point spread function in the RM spectrum (called the RM spread function or RMSF). 
By default, it assumes that the source emits with a spectral index $\alpha=0$. 
Because synchrotron sources often have non-zero spectral indices, this leads to a mismatch between the real emission from the source and what RMClean assumes for this emission; as a result, RMClean has to introduce additional spurious components to explain the observations.

In SL17 we developed a mathematical framework that includes both spectral index effects and a variation in sensitivity across the observing band; we applied this framework to observations of pulsars in the Galactic Centre \citep{schnitzeler2016}.
We showed that RM synthesis maximizes the likelihood only under certain conditions that are often not met by actual radio sources (e.g., the requirement that their polarized flux density spectral index is zero).
We also showed that the practice of dividing the measured linear Stokes parameters $Q$ and $U$ by Stokes $I$, which is often used to correct for spectral index effects, in combination with RM synthesis does not maximize the likelihood.
Furthermore, a single pixel can contain sources with different spectral indices in Stokes $I$, and different parts of a radio source can emit different polarization fractions (for example, the cores of active galactic nuclei are often heavily depolarized but bright in Stokes $I$, whereas the polarization fraction is higher further from the core).
Then it also makes little physical sense to divide Stokes $Q$ and $U$ by Stokes $I$.
Here we extend the mathematical framework from SL17, combining analytical and numerical methods, so that also complex sources like active galactic nuclei and the interstellar medium of the Milky Way or nearby galaxies can be analysed. 
In this paper we consider single lines of sight, like we did in S17, but contrary to that paper, we will assume that the correct model for the radio source is not known a priori, but has to be selected from a list of available models.

In Section~\ref{parameters.sec}, we describe which models we fit to the data, how we fit these models, and which assumptions we make. 
A comparison between the equations we derive, and the equations for RM synthesis, is a powerful tool for identifying issues with these techniques. We present this analysis in Section~\ref{rayleigh.sec}. There we also quantify the resolution in RM synthesis and QU fitting. 
Then, in Section~\ref{firestarter.sec}, we introduce the software tool we developed for fitting and ranking source models using a number of different metrics. 
We test this software on simulated observations of different types of sources in Section~\ref{spass.sec}, and we analyse its strengths and weaknesses. Also, we analyse the impact of noise and of the resolution in RM on our results.
In Section~\ref{related_topics.sec} we discuss a number of topics related to QU fitting and RM synthesis/RMClean. 
We summarize our results in Section~\ref{conclusions.sec}.

\section{Parameter estimation}\label{parameters.sec}
To fit source models to actual observations we need to describe the observing setup and the properties of the noise.
We assume that Stokes $Q$ and $U$ flux densities have been measured for $N_\mathrm{ch}$  frequency channels, which are statistically independent (there is no correlation between the noise in different channels).
Furthermore, the flux densities in $Q$ and $U$ do not show any offsets from zero across the frequency band.
We indicate these observations with $Q_{\mathrm{obs},i}$ and $U_{\mathrm{obs},i}$, where `$i$' is the channel index (the index `$i$' retains this meaning throughout our paper). 
The noise in each frequency channel follows a Gaussian distribution with zero mean and a variance $\sigma_{Q,i}^2$ or $\sigma_{U,i}^2$. 
These variances can be different for the two Stokes parameters, and are allowed to vary between channels.
For this observing setup the log likelihood is
\begin{eqnarray}
\lefteqn{
\log{\Lambda} =
} \label{loglikelihood.eqn} \nonumber \\
& & -\frac{1}{2}\sum_{i=1}^{N_\mathrm{ch}}
          \left[\left(\frac{Q_{\mathrm{obs},i} - Q_{\mathrm{mod},i}}{\eta\,\sigma_{Q,i}}\right)^2 + \left(\frac{U_{\mathrm{obs},i} - U_{\mathrm{mod},i}}{\eta\,\sigma_{U,i}}\right)^2 \right] \nonumber \\
& & - \sum_{i=1}^{N_\mathrm{ch}} \left[\log\left(\sigma_{Q,i}\right) + \log\left(\sigma_{U,i}\right)\right] 
-N_\mathrm{ch}\left[\log\left(2\upi\right)+2\log\left(\eta\right)\right]\, , \nonumber \\
 & & 
\end{eqnarray}
where $Q_{\mathrm{mod},i}$  and $U_{\mathrm{mod},i}$  indicate the modelled Stokes $Q$ and $U$ flux densities in each channel and $\eta$ is a multiplicative scale factor for the measured noise variances in Stokes $Q$ and $U$. 
We will assume that these noise variances are correct, and we set $\eta=1$.
Furthermore, we assume that Faraday rotation across individual frequency channels is small, so that we can approximate the net (integrated) polarization vector of a channel with the polarization vector at the mean wavelength squared of that channel (see also \citealt{schnitzeler2015}).
We will write the linear polarization vector as $\bmath{L} = Q+\mathrm{i}U$.

We assume that each part of the source emits a power-law spectrum in Stokes $I$ and $L$ as a function of frequency (e.g., \citealt{ginzburg1965}), and we will include the flux density spectral index $\alpha$ as a separate parameter that has to be fitted. 
Deviations from a power law in Stokes $I$ point to additional physical processes that should be included when interpreting the data (e.g., a spectral turnover due to synchrotron self-absorption or free-free absorption, or a spectral break due to ageing of the cosmic ray population).
Since we assume that the intrinsic emission spectrum follows a power law, deviations between the measurements and a power law can be modelled as being due to depolarization by magnetic fields inside or in front of the source.
Also, by fitting for the spectral index when modelling radio sources, we avoid using ratios of flux densities $Q/I$ and $U/I$, which we demonstrated in SL17 does not maximize the likelihood.

In this Section we first give an overview of the different physical models that we consider in our analysis (Section~\ref{continuous_model.sec}).
Then, in Section~\ref{maxl.sec}, we derive a matrix equation for calculating the maximum-likelihood estimators for Stokes $Q$ and $U$ of each source component, at a chosen reference frequency.
This equation lets us focus our computing resources on numerically searching only over the non-linear model parameters.

\subsection{Overview of models}\label{continuous_model.sec}
\begin{figure}
\begin{centering}
\resizebox{0.6\hsize}{!}{\includegraphics{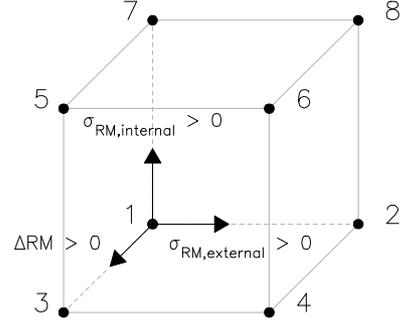}}
\caption{Grid of source models that we introduce in Section~\ref{continuous_model.sec}.
}
\label{modelgrid.fig}
\end{centering}
\end{figure}

We classify models based on how the emitted polarized flux density $|\bmath{L}|$ varies with RM: polarization observations let us distinguish only between models with different distributions of $\bmath{L}(\mathrm{RM})$.
These models were discussed by, e.g., \cite{burn1966}, \cite{sokoloff1998}, and \cite{schnitzeler2015b}. 
Fig.~\ref{modelgrid.fig} gives an overview of the models we consider, highlighting relations between models.

\noindent (1) Point source in RM. The modelled Stokes $Q$ and $U$ flux density in frequency channel $i$ is
\begin{equation}
\begin{pmatrix}
Q_{\mathrm{mod},i} \\
U_{\mathrm{mod},i}
\end{pmatrix}
=
\begin{pmatrix}
c_{i} & -s_{i} \\
s_{i} & c_{i} \\
\end{pmatrix}
\begin{pmatrix}
Q_{\mathrm{ref}} \\
U_{\mathrm{ref}}
\end{pmatrix}
\left(\frac{\nu_i}{\nu_\mathrm{ref}}\right)^{\alpha}\, .
\label{QUmodel_single_peaks.eqn} 
\end{equation}
`ref' refers to the reference frequency of the power law, and $c_{i}$ and $s_{i}$ are abbreviations for $\cos{\left[2\mathrm{RM_0}\left(c/\nu_i\right)^2\right]}$ and $\sin{\left[2\mathrm{RM_0}\left(c/\nu_i\right)^2\right]}$, respectively, where RM$_0$ is the RM of the foreground, $c$ the speed of light, and $\nu_i$ the frequency of the channel. The matrix in equation~(\ref{QUmodel_single_peaks.eqn}) describes Faraday rotation of the polarized emission. Alternatively, equation~(\ref{QUmodel_single_peaks.eqn}) can be expressed using complex numbers, allowing for a more compact notation:
\begin{eqnarray}
\bmath{L}_{\mathrm{mod},i} = \bmath{L}_\mathrm{ref}\,\left(\frac{\nu_i}{\nu_\mathrm{ref}}\right)^{\alpha}\, \mathrm{exp}\left[2\mathrm{i}\mathrm{RM_0}\left(c/\nu_i\right)^2\right]\, . 
\label{point_source.eqn}
\end{eqnarray}
In SL17 we developed a maximum likelihood (ML)-based formalism for determining the parameters of this type of source.
\\
(2) Gaussian $|\bmath{L}(\mathrm{RM})|$ distribution, which occurs if a turbulent foreground screen with a Gaussian RM distribution (mean RM$_0$ and variance $\sigma_\mathrm{RM,external}^2$) lies in front of a uniformly emitting source. 
If the telescope beam encompasses many turbulent cells, we observe a polarized flux density
\begin{eqnarray}
\lefteqn{
\bmath{L}_{\mathrm{mod},i}
 = 
 \bmath{L}_\mathrm{ref}\, \left(\frac{\nu_i}{\nu_\mathrm{ref}}\right)^\alpha
 \times
 }
 \nonumber \\
 & & 
  \mathrm{exp}\left[-2\sigma_\mathrm{RM,external}^2\left(c/\nu_i\right)^4+2\mathrm{i}\mathrm{RM_0}\left(c/\nu_i\right)^2\right]\, .
\label{gaussian_screen.eqn}
\end{eqnarray}
A 2D Gaussian source that lies behind a linear gradient in RM shows similar behaviour:
\begin{eqnarray}
\lefteqn{
\bmath{L}_{\mathrm{mod},i} 
 = 
 \bmath{L}_\mathrm{ref}\, \left(\frac{\nu_i}{\nu_\mathrm{ref}}\right)^\alpha
 \times
 }
 \nonumber \\
 & & 
 \mathrm{exp}\left[-2\Delta\mathrm{RM}^2\left(c/\nu_i\right)^4+2\mathrm{i}\mathrm{RM_0}\left(c/\nu_i\right)^2\right]\, , 
\label{gaussian_source.eqn}
\end{eqnarray}
where 
\begin{eqnarray}
\Delta\mathrm{RM}^2 & = & \left(\frac{\upartial\mathrm{RM}}{\upartial  x}\sigma_x\right)^2 + \left(\frac{\upartial\mathrm{RM}}{\upartial  y}\sigma_y\right)^2 \nonumber
\end{eqnarray}
quantifies the change in RM across the source which measures $\sigma_x \times \sigma_y$ standard deviations in the orthogonal directions `$x$' and `$y$'. 
In this case RM$_\mathrm{0}$, the RM of the foreground medium, is equal to the RM at the centre of the source.
\\
(3) Rectangular $|\bmath{L}(\mathrm{RM})|$ distribution, which occurs if RM increases linearly inside a uniformly emitting source (`Burn slab') or if a uniformly emitting source lies behind a linear gradient in RM:
\begin{eqnarray}
\lefteqn{
\bmath{L}_{\mathrm{mod},i} 
 = 
 \bmath{L}_\mathrm{ref}\, \left(\frac{\nu_i}{\nu_\mathrm{ref}}\right)^\alpha
 \mathrm{sinc}\left[\Delta\mathrm{RM}\left(c/\nu_i\right)^2\right]
 \times
 }
 \nonumber \\
 & &
 \mathrm{exp}\left[2\mathrm{i}\mathrm{RM_c}\left(c/\nu_i\right)^2\right]\, , 
\label{burn_slab.eqn}
\end{eqnarray}
where RM$_\mathrm{c}$ is the mean RM of the emission, and $\Delta$RM the change in RM across the source.
\\
(4) Model (3) when it is depolarized by the turbulent foreground screen from model (2).
The behaviour of the polarization vector is described by the product of equation~(\ref{burn_slab.eqn}) and exp$\left[-2\sigma^2_\mathrm{RM,external}\left(c/\nu\right)^4\right]$.
\\
(5) A uniformly emitting source with an embedded large-scale and small-scale magnetic field. If there are many turbulent cells within the telescope beam, then we observe
\begin{eqnarray}
\bmath{L}_{\mathrm{mod},i}
=
\bmath{L}_\mathrm{ref}\,
\left(\frac{\nu_i}{\nu_\mathrm{ref}}\right)^\alpha \frac{1-\mathrm{e}^{-S_i}}{S_i} \mathrm{exp}\left[2\mathrm{i}\mathrm{RM}_0\left(c/\nu_i\right)^2\right]\, ,  
\label{ifd.eqn}
\end{eqnarray}
where $S_i= 2\sigma_\mathrm{RM,internal}^2\left(c/\nu_i\right)^4 - 2\mathrm{i}\Delta\mathrm{RM}\left(c/\nu_i\right)^2$. 
RM$_0$ is the RM of the foreground, $\sigma_\mathrm{RM,internal}^2$ the variance of the Gaussian RM distribution inside the source, and $\Delta$RM the change in RM over the source region due to the large-scale magnetic field inside the source.
\\
(6) Internal Faraday dispersion which is depolarized by the turbulent foreground screen from model (2).
\\
(7) Internal Faraday dispersion without a large-scale magnetic field present ($\Delta$RM = 0), and 
\\
(8) Model (7) combined with model (2). 
\\
In Section~\ref{spass.sec} we explain why it is necessary to test not only model 6, the most complex model that we consider, but also each of its parent models (which have at least one component less than their daughter).

\cite{horellou2014} showed that with proper modifications some of the above models also describe helical magnetic fields inside or outside the source region.

The telescope beam can encompass more than one of these source types, or there could be several regions with Faraday rotation and synchrotron emission along the line of sight.
The observed polarization vector is then equal to the vector sum of the contributions by the individual sources; if sources lie behind one another then the foreground source can induce additional Faraday rotation and/or depolarization of the emission from the background source.

\subsection{Maximizing the likelihood}\label{maxl.sec}
The polarization behaviour of models 1-4 can be written as 
\begin{eqnarray}
\bmath{L}\left(\nu\right) = \bmath{L}_\mathrm{ref}\, 
\mathrm{fnc}\left(\nu\right)\mathrm{exp}\left[2\mathrm{i}\mathrm{RM}\left(c/\nu_i\right)^2\right]\, , 
\label{polvecone.eqn}
\end{eqnarray}
where fnc$\left(\nu\right)$ is a real-valued function that describes the amplitude modulation of the signal. 
Models 5-8 can be rewritten in a form similar to equation~(\ref{polvecone.eqn}) by multiplying the numerator and denominator in these models by the complex conjugate of $S_i$; the resulting product $S_i^*\left(1-\mathrm{e}^{-S_i}\right)$ in the numerator can then be expanded to the following sum of four functions: 
\begin{eqnarray}
\bmath{L}\left(\nu\right) = \bmath{L}_\mathrm{ref}
\sum_{l=1}^4 \mathrm{fnc}_l\left(\nu\right)\mathrm{exp}\left[2\mathrm{i}\left(\mathrm{RM}_l\left(c/\nu\right)^2 + \Delta\chi_{0,l}\right)\right]\, ,
\label{polvectwo.eqn}
\end{eqnarray}
where
\begin{eqnarray}
\mathrm{fnc}_1\left(\nu\right) & = & 2\sigma_{\mathrm{RM,internal}}^2\left(c/\nu\right)^4 
\left(\nu/\nu_\mathrm{ref}\right)^\alpha
/\left( S_i S_i^*\right) \nonumber \\
\mathrm{fnc}_2\left(\nu\right) & = & 2\Delta\mathrm{RM}\left(c/\nu\right)^2 
\left(\nu/\nu_\mathrm{ref}\right)^\alpha
/\left( S_i S_i^*\right) \nonumber \\
\mathrm{fnc}_3\left(\nu\right) & = & - \mathrm{fnc}_1\left(\nu\right)\,\mathrm{e}^{-2\sigma_{\mathrm{RM,internal}}^2\left(c/\nu\right)^4} \nonumber \\
\mathrm{fnc}_4\left(\nu\right) & = & - \mathrm{fnc}_2\left(\nu\right)\, \mathrm{e}^{-2\sigma_{\mathrm{RM,internal}}^2\left(c/\nu\right)^4} \nonumber \\
\mathrm{RM}_l & = &
\begin{cases}
\mathrm{RM}_0 \hfill (l=1,2) \nonumber \\
\mathrm{RM}_0 + \Delta\mathrm{RM} \hfill (l=3,4) \nonumber 
\end{cases}\\
\Delta\chi_{0,l} & = & 
\begin{cases}
0 \hfill (l=1,3) \nonumber \\
\upi/4 \hfill (l=2,4) 
\end{cases}
\end{eqnarray}
`$^*$' indicates complex conjugation. In models 6 and 8 all $\mathrm{fnc}_l$ should be multiplied by $\mathrm{exp}\left[-2\sigma_{\mathrm{RM,external}}^2\left(c/\nu\right)^4\right]$).
By writing models 5-8 this way, we can determine $Q_{\mathrm{ref}}$ and $U_{\mathrm{ref}}$ in the same way in all models. This simplifies the computer program we will introduce in Section~\ref{firestarter.sec}.

The likelihood can be maximized for the linear model parameters by taking the partial derivative of equation~\ref{loglikelihood.eqn} with respect to the $Q_{\mathrm{ref},j}$ and $U_{\mathrm{ref},j}$ of each source component $j$, and setting the result equal to zero.
If there are $N_\mathrm{cmp}$ source components ($j=1,..,N_\mathrm{cmp}$), then this leads to the following equations:
\begin{eqnarray}
\begin{pmatrix}
\mathbfss{A}_{1,1} & \ldots & \mathbfss{A}_{1,N_\mathrm{cmp}} \\
\vdots & \ddots & \vdots \\
\mathbfss{A}_{N_\mathrm{cmp},1} & \ldots & \mathbfss{A}_{N_\mathrm{cmp},N_\mathrm{cmp}} \\
\end{pmatrix}
\begin{pmatrix}
\begin{pmatrix}
Q_{\mathrm{ref},1}\\
U_{\mathrm{ref},1}\\
\end{pmatrix} \\
\vdots \\
\begin{pmatrix}
Q_{\mathrm{ref},N_\mathrm{cmp}}\\
U_{\mathrm{ref},N_\mathrm{cmp}}\\
\end{pmatrix}\\
\end{pmatrix} = 
\begin{pmatrix}
\bmath{E_1} \\
\vdots \\
\bmath{E_{N_\mathrm{cmp}}}
\end{pmatrix}
\label{matrix.twocomp.eqn}
\end{eqnarray}
where 
\begin{eqnarray}
\mathbfss{A}_{j1,j2} & = &
\begin{pmatrix}
B_{j1,j2} & C_{j1,j2} \nonumber \\
C_{j2,j1} & D_{j1,j2} \nonumber \\
\end{pmatrix} \\
B_{j1,j2} & = & 
\sum_{i=1}^{N_\mathrm{ch}} \left(\frac{cf_{j1,i}cf_{j2,i}}{\sigma_{Q,i}^2}+\frac{sf_{j1,i}sf_{j2,i}}{\sigma_{U,i}^2}\right) \nonumber\\
C_{j1,j2} & = & 
\sum_{i=1}^{N_\mathrm{ch}} \left(-\frac{cf_{j1,i}sf_{j2,i}}{\sigma_{Q,i}^2}+\frac{sf_{j1,i}cf_{j2,i}}{\sigma_{U,i}^2}\right) \nonumber\\
D_{j1,j2} & = & 
\sum_{i=1}^{N_\mathrm{ch}} \left(\frac{sf_{j1,i}sf_{j2,i}}{\sigma_{Q,i}^2}+\frac{cf_{j1,i}cf_{2,i}}{\sigma_{U,i}^2}\right)  \nonumber\\
\bmath{E_j} & = &
\sum_{i=1}^{N_\mathrm{ch}}
\begin{pmatrix}
cf_{j,i} & sf_{j,i} \\
-sf_{j,i} & cf_{j,i} \\
\end{pmatrix}
\begin{pmatrix}
Q_{\mathrm{obs},i}/\sigma_{Q,i}^2 \\
U_{\mathrm{obs},i}/\sigma_{U,i}^2 \\
\end{pmatrix}
 \nonumber \\
cf_{j,i} & = & \sum_{l=1}^4 c_{j,l,i}\, f_{j,l,i}\nonumber \\
sf_{j,i} & = & \sum_{l=1}^4 s_{j,l,i}\, f_{j,l,i} \nonumber \\
f_{j,l,i} & = &  \mathrm{fnc}_{j,l}\left(\nu_i\right)\mathrm{exp}\left[2\mathrm{i}\left(\mathrm{RM}_{j,l}\left(c/\nu_i\right)^2 + \Delta\chi_{0,j,l}\right)\right] \, . \nonumber
\end{eqnarray}
`$c_{j,l,i}$' and `$s_{j,l,i}$' refer to the values of $c_{i}$ and $s_{i}$ of model component $j$, and the index $l$ is defined by equation~(\ref{polvectwo.eqn}).
For models 1-4, $l$ takes on only one value, therefore $cf_{j,i}$ and $sf_{j,i}$ simplify to $c_{i}$ or $s_{i}$ times fnc$\left(\nu\right)$ from equation~(\ref{polvecone.eqn}).

The likelihood can now be calculated by specifying only the non-linear parameters in the model under investigation, then using equation~(\ref{matrix.twocomp.eqn}) to find the values for $Q_{\mathrm{ref},j}$ and $U_{\mathrm{ref},j}$ that maximize the likelihood.
Since numerical techniques for maximizing the likelihood need to take into account fewer parameters, reducing the dimensionality of the search space, processing is sped up.

\section{Comparison between QU fitting and RM synthesis; the Rayleigh criterion}\label{rayleigh.sec}
Writing the expression for RM synthesis in matrix form,
\begin{eqnarray}
\begin{pmatrix} Q_\mathrm{ref} \\ U_\mathrm{ref} \\
\end{pmatrix} = 
\frac{1}{N_\mathrm{ch}}\sum_{i=1}^{N_\mathrm{ch}}
\begin{pmatrix}
c_i & s_i \\
-s_i & c_i\\
\end{pmatrix}
\begin{pmatrix} Q_{\mathrm{obs},i} \\ U_{\mathrm{obs},i} \\
\end{pmatrix}\, , 
\label{rmsynthesis.eqn}
\end{eqnarray}
and comparing this equation with equation~(\ref{matrix.twocomp.eqn}), it becomes clear that RM synthesis and QU fitting are connected.
Each $\bmath{E_j}$ in equation~(\ref{matrix.twocomp.eqn}) expresses how polarization vectors measured at different frequencies are weighted, aligned (derotated), and summed.
The matrix on the left-hand side of equation~(\ref{matrix.twocomp.eqn}) ensures that the result of this calculation is normalized correctly.
If we are modelling a single point source that has $\alpha_j=0$, and if the noise variances in Stokes $Q$ and $U$ are equal, $\sigma_{Q,i}^2=\sigma_{U,i}^2\equiv\sigma_{L,i}^2$, then equation~(\ref{matrix.twocomp.eqn}) simplifies to the equation for weighted RM synthesis, as we showed in section~2.3.1 from SL17.
If there are two point sources then each matrix $\mathbfss{A}_{j,j}$ is a diagonal matrix, all elements on the diagonal being equal to $\sum_{i=1}^{N_\mathrm{ch}} 1/\sigma_{L,i}^2\,(\nu_i/\nu_\mathrm{ref})^{2\alpha_j}$. 
The matrix $\mathbfss{A}_{j1,j2}$ can be rewritten using the double-angle formulae from trigonometry, leading to 
\begin{eqnarray}
\lefteqn{
\bmath{E_1} = 
\sum_{i=1}^{N_\mathrm{ch}} \frac{1}{\sigma_{L,i}^2}\left(\frac{\nu_i}{\nu_\mathrm{ref}}\right)^{\alpha_1}
\left[
\begin{pmatrix}
Q_{\mathrm{ref},1} \\
U_{\mathrm{ref},1}
\end{pmatrix}
\left(\frac{\nu_i}{\nu_\mathrm{ref}}\right)^{\alpha_1} + \right. } 
\nonumber \\
& &  \left.
\begin{pmatrix}
\cos\left[\hspace{2mm}\right] & -\sin\left[\hspace{2mm}\right] \\
\sin\left[\hspace{2mm}\right] & \cos\left[\hspace{2mm}\right] \\
\end{pmatrix}
\begin{pmatrix}
Q_{\mathrm{ref},2} \\
U_{\mathrm{ref},2}
\end{pmatrix}
\left(\frac{\nu_i}{\nu_\mathrm{ref}}\right)^{\alpha_2}
\right]\, ,  
\label{beat.eqn}
\end{eqnarray}
where $\cos\left[\hspace{2mm}\right]=\cos\left(2\Delta\mathrm{RM}\lambda_i^2\right)$, $\sin\left[\hspace{2mm}\right]=\sin\left(2\Delta\mathrm{RM}\lambda_i^2\right)$, and $\Delta$RM$\equiv$RM$_2$-RM$_1$ (a similar expression holds for $\bmath{E_2}$).

Equation~(\ref{beat.eqn}) would be identical to the equation for weighted RM synthesis if the contribution from the second source to $\bmath{E_1}$ is zero. 
To quantify when this happens, assume that each source has a spectral index of zero, meaning that the emitted polarization vectors all have the same length across the frequency band, and that we are observing not in frequency but in wavelength squared.
Then this first `null' occurs when the polarization vectors emitted by source two change their orientation by $\upi$ radians across the band, i.e., $\Delta\lambda^2\,\Delta\mathrm{RM} = \upi$, or 
\begin{eqnarray}
\Delta\mathrm{RM}_\mathrm{Rayleigh} = \frac{\upi}{\Delta\lambda^2} \approx 0.83\, \mathrm{FWHM(RMSF)}\, ,
\label{rayleigh.eqn}
\end{eqnarray}
where $\Delta\lambda^2\equiv\lambda^2_\mathrm{max}-\lambda^2_\mathrm{min}$ is the wavelength squared coverage of the observations (we used equation~6 from \citealt{schnitzeler2009} to quantify the full width at half-maximum (FWHM) of the RM spread function: FWHM (\radm) = 3.8/$\Delta\lambda^2$).
In optics, the Rayleigh criterion provides a rough estimate for when two point sources can be easily resolved: this occurs when the second source lies at the first null of the diffraction pattern produced by the first source. 
The same idea underlies equation~(\ref{rayleigh.eqn}), therefore we added a subscript `Rayleigh' to this equation.
The second part of equation~(\ref{beat.eqn}), which quantifies the contribution by source two to $\bmath{E_1}$, is identical to the expression for the RMSF.
Therefore, equation~(\ref{rayleigh.eqn}) also describes the resolution of RM spectra, and we conclude that the resolution in QU fitting and RM synthesis is the same.

Of course, we observe at regular intervals in frequency, not wavelength squared\footnote{In \cite{schnitzeler2015} we discuss differences between these observing setups, the effect of channel weighting functions, and their impact on RM synthesis in more detail.}, and as a result of this the polarization vectors emitted by the source across the frequency band will not cancel completely at $\Delta\mathrm{RM}=\Delta\mathrm{RM}_\mathrm{Rayleigh}$.
Furthermore, the polarization vectors will also not cancel completely at $\Delta\mathrm{RM}_\mathrm{Rayleigh}$ if the spectral index of the source is not equal to zero.
This is illustrated by the solid and dashed lines in Fig.~\ref{steepsp.fig}.
Therefore, equation~(\ref{rayleigh.eqn}) should be considered as a crude way for estimating whether two point sources can be resolved; techniques developed in information theory or Bayesian statistics provide an alternative approach for determining whether a simple model describes the data better than a more complex model
(\citealt{marti-vidal2012} used similar reasoning to determine if sources are resolved in radio interferometric observations).
We explore this possibility when we consider automated model selection and ranking in the next sections.

A comparison between equation~(\ref{matrix.twocomp.eqn}) and the equation for RM synthesis, equation~(\ref{rmsynthesis.eqn}), highlights an important difference between QU fitting and RM synthesis: RM synthesis does not correctly take into account how sources with comparable RMs interfere with each another.
To simplify the comparison, assume that all sources have spectral indices of zero, and that the noise variances in Stokes $Q$ and $U$ are equal and constant across the frequency band.
If we are looking for two point sources, then we can find $Q_\mathrm{ref}$ and $U_\mathrm{ref}$ of each source from equation~(\ref{matrix.twocomp.eqn}) if we specify the RM of each source.
Calculating the RM spectrum at the same two RMs can be written as a matrix equation similar to equation~(\ref{matrix.twocomp.eqn}), resulting in the expression for $\left((Q_{\mathrm{ref},1},U_{\mathrm{ref},1}),(Q_{\mathrm{ref},2},U_{\mathrm{ref},2}) \right)^\mathrm{T}$.
However, in the case of RM synthesis, the equivalent of the matrix on the left-hand side of equation~(\ref{matrix.twocomp.eqn}) is a diagonal matrix, all $\mathbfss{A}_{j1,j2}$ ($j1 \ne j2$) being zero.
Earlier in this section we showed that the matrices $\mathbfss{A}_{j1,j2}$ describe how sources $j1$ and $j2$ influence each other: in an RM spectrum this is visible as two overlapping RMSFs.
If the matrices $\mathbfss{A}_{j1,j2}$ are all zero, then this means that RM synthesis does not account for this interference in a way that maximizes the likelihood.
Alternatively, one could argue that RM synthesis only approximates maximizing the likelihood if the sources emit at very different RMs, $|\Delta\mathrm{RM}| \gg |\Delta\mathrm{RM}_\mathrm{Rayleigh}|$: in that case the amplitude of the RMSF centred on the first source has dropped to zero at the RM of the second source, and vice versa, so that the two sources do not interfere.
If this condition does not hold, then this can lead to artefacts in RM synthesis and RMClean.
QU fitting does not suffer from this.

\section{Model selection; the $\textsc{Firestarter}$ program}\label{firestarter.sec}
We wrote a program called $\textsc{Firestarter}$ to fit models to measurements of Stokes $Q$ and $U$ as a function of frequency, and to rank these models based on the quality of the fit and the number of parameters in each model.
$\textsc{Firestarter}$ incorporates information on the noise variances of these parameters, and builds on results from previous sections. 
The program can be downloaded from the following URL\footnote{https://github.com/dschnitzeler/firestarter}.
Here we describe how $\textsc{Firestarter}$ works, while in Section~\ref{spass.sec} we test $\textsc{Firestarter}$ under different conditions, using Monte Carlo simulations. 

The basic layout of $\textsc{Firestarter}$ is as follows:
\begin{enumerate}
\renewcommand{\theenumi}{(\arabic{enumi})}
\item Compile a list of models which should be tested. Repeat the following steps for each model in this list, but skip all daughter models if a parent model is flagged during the fitting process.
\item 
The program relies on the Levenberg-Marquardt (LM) algorithm (\citealt{levenberg1944}, \citealt{marquardt1963}, \citealt{more1978}, \citealt{markwardt2009}) to fit a model to the data\footnote{We assumed that in our observations the frequency channels are statistically independent, and follow Gaussian probability density functions. In this case maximizing the likelihood is identical to minimizing $\chi^2$, for which the LM algorithm was developed.}. 
To define a starting point for the LM algorithm, the best-fitting parameters for the parent model are combined with reasonable estimates for the parameters that describe the new model component.
The program finds an initial value for the RM of this new component by applying RM synthesis to the residual Stokes $Q$ and $U$ spectra (original measurements of these Stokes parameters minus the prediction for the best-fitting parent model).
The RM associated with the highest polarized flux density in this RM spectrum, together with values of -0.7 for the spectral index and 1/4th the value of FWHM(RMSF) for all other components, then form the initial model parameters for the new source component.
If the peak polarized flux density in the residual RM spectrum has a signal-to-noise ratio below 1.5, this model and all its daughters are skipped (see Section~\ref{rmcleanstop.sec}).
\item 
Run the LM algorithm to find the best-fitting model parameters. 
The parameters of both the parent model and of the new component are allowed to vary in order to produce the best fit to the data. 
The LM algorithm provides uncertainties for the fitted non-linear model parameters; we determine the uncertainties in $Q_\mathrm{ref}$ and $U_\mathrm{ref}$ of each model component by fitting parabolas to the log likelihood (e.g., \citealt{avni1976}, SL17).
\item Calculate metrics that measure the quality of the fit.
\end{enumerate}
At the core of this program the LM algorithm fits the non-linear model parameters; we use equation~(\ref{matrix.twocomp.eqn}) to calculate the ML estimators for $Q_\mathrm{ref}$ and $U_\mathrm{ref}$. 
Instead of fitting all non-linear model components simultaneously, we fit models with multiple components in several iterations, adding one additional component in each iteration (we learned this trick from $\textsc{DIFMAP}$, \citealt{shepherd1997}). 
This approach makes it easier to find starting values for the LM algorithm, and it reduces the risk of converging to a local maximum in the likelihood.

To create a list of models for step (1), $\textsc{Firestarter}$ uses as input a selection of model components from Section~\ref{continuous_model.sec} that should be tested, together with a specification of the maximum number of allowed components.
Without additional information, the number and order of the model components might not be clear a priori, in which case it is important to consider all possible combinations of model components.
By fitting models with any number of components up to the specified maximum, the program can test if a simpler model describes the data better than a more complex model.
Resolving a source on the sky can give vital clues about which components should be included, and which components can be ruled out, for example, linear gradients in front of a resolved source.

In step~(3), $\sigma_\mathrm{RM, internal}$, $\sigma_\mathrm{RM, external}$, and $\Delta\mathrm{RM}$ are restricted to values $\ge 0$.
The factor $S_i S_i^*$ in the denominator of equation~(\ref{polvectwo.eqn}) can become very small during model fitting, leading to numerical instabilities. To prevent this, we set $\sigma_{\mathrm{RM,internal}}$, $\Delta\mathrm{RM}$, and their associated uncertainties to zero if $S_i S_i^* < $ ten times the machine precision.
The program evaluates the product $S_i S_i^*$ at the highest observed frequency, hence the smallest $\lambda^2$ covered by the observations.
Only spectral indices between -6 and +3 are allowed in the fits, which covers the spectral indices of all known pulsars and most active galactic nuclei (\citealt{lorimer1995}, \citealt{bates2013}).

$\textsc{Firestarter}$ automatically ranks models based on how well they describe the data, relying on concepts from the fields of information theory and statistics.
After finding the best-fitting model parameters it calculates four metrics:
\begin{enumerate}
\renewcommand{\theenumi}{(\arabic{enumi})}
\item The log likelihood ratio, which quantifies the significance of the detection (e.g., \citealt{wilks1938})
\item The reduced $\chi^2$ of the fit
\item The Akaike Information Criterion (AIC; \citealt{akaike1973, akaike1974})
\item The Bayesian Information Criterion (BIC; \citealt{schwarz1978}, \citealt{raftery1995})
\end{enumerate}
A model is excluded from this ranking if it has been flagged during the fitting process.
The AIC and BIC quantify in different ways which model describes the data best: AIC measures the Kullback-Leibler divergence between the observations and the model, while the difference in BIC values of two models approximates the log of the Bayes factor of those models. 
The AIC and BIC are calculated from
\begin{eqnarray}
\mathrm{AIC} & = & -2\log\left(\Lambda_\mathrm{max}\right) + 2\,\Delta N_\mathrm{par} \label{AIC.eqn} \\
\mathrm{BIC} & = & -2\log\left(\Lambda_\mathrm{max}\right) + \Delta N_\mathrm{par}\,\log\left(2\,N_\mathrm{ch}\right) \label{BIC.eqn}\, , 
\end{eqnarray}
where $\log\left(\Lambda_\mathrm{max}\right)$ is the maximum in log likelihood of the model fit, and $\Delta N_\mathrm{par}$ = $N_\mathrm{par}$ - $N_\mathrm{par,0}$, the number of parameters of the fitted model minus the number of parameters in the null model (which contains no parameters but can include, for example, offsets or the ML estimator for $\eta$).
In our case the null model has no free parameters, therefore $\Delta N_\mathrm{par}$ equals the number of parameters in each model. 
For the AIC, BIC, and the reduced $\chi^2$ of the fit, smaller is better.
If the number of observations is small compared to the number of model parameters ($N_\mathrm{ch}/N_\mathrm{par}\, \la\, 20$ for a linear regression model, \citealt{burnham2004}) then a bias correction term has to be applied to the AIC (\citealt{sugiura1978}, \citealt{hurvich1989, hurvich1995}). 
This correction term has been derived for linear models and in a few other cases, but as far as we know not for the non-linear models we consider; therefore we cannot apply the small-sample correction.
Both the AIC and BIC only provide a relative ranking of models. By calculating the reduced $\chi^2$ of each model we quantify the quality of each fit in an absolute sense.

The difference $\Delta_j=\mathrm{AIC}_j-\mathrm{AIC}_\mathrm{min}$ between the AIC value of model $j$ and the model with the smallest AIC quantifies whether the data prefer the model with the lowest AIC to model $j$ (the BIC can be used instead of the AIC to calculate $\Delta_j$).
A value $\Delta_j > 10$ indicates a very strong preference, whereas if $\Delta_j < 2$ the preference is weak (\citealt{raftery1995} and \citealt{burnham2004}). 
Raftery translates these values of $\Delta_j$ into Bayes factors of approximately 150 and 3, respectively.
This coarse selection based on $\Delta_j$ can be refined by using $\Delta_j$ to calculate weights for each of the tested models:
\begin{eqnarray}
w_j \approx \frac{\exp\left(-\Delta_j/2 \right)}{\sum_{k=1}^{N_\mathrm{mod}} \exp\left(-\Delta_k/2 \right)}\, , 
\label{model_weights.eqn}
\end{eqnarray}
where the index $k$ loops over all models being tested.
$w_j$ quantifies the probability that the model with index $j$ generated the data \citep{wassermann2000}.
This suggests a connection with Bayesian model selection, as discussed by \cite{raftery1995} and in section~4 of \cite{burnham2004}.
Because the data are noisy, it is possible that several models have non-zero weights $w_j$, meaning that they can all explain the observations up to some degree.
This is not taken into account if one uses only the model with the lowest AIC or BIC value in data analysis.
As suggested by Raftery and Burnham \& Anderson, one can use the weights $w_j$ to calculate averages of parameters and their associated uncertainties (this is known as `model averaging'):
\begin{eqnarray}
\bar{\theta} & \approx & \sum_{k=1}^{N_\mathrm{mod}} w_j \hat{\theta}_j \Big / \sum_{k=1}^{N_\mathrm{mod}} w_j \\
\mathrm{err}\left(\bar{\theta}\right) & \approx & 
\sum_{k=1}^{N_\mathrm{mod}} w_j 
\sqrt{\mathrm{var}\left(\hat{\theta}_j\right) + \left(\hat{\theta}_j - \bar{\theta}\right)^2} \Big / \sum_{k=1}^{N_\mathrm{mod}} w_j \, ,
\end{eqnarray}
where $\hat{\theta}_j$ indicates the ML estimator for parameter $\theta$ in model $j$, and $\mathrm{var}\left(\hat{\theta}_j\right)$ is the variance of this estimator.
If the parameter of interest is absent from a model then $w_j = 0$ for that model.

We consider model fits to the data to be reliable if their reduced $\chi^2 < 1 + 5\sqrt{2/N_\mathrm{dof}}$, and if the fitted $\alpha$ are neither -6 or +3 (which means that the LM algorithm converged to the minimum or maximum allowed value). Only these fits are included when calculating model weights $w_j$.
$N_\mathrm{dof}$ is the number of degrees of freedom of the fit, which is equal to $2N_\mathrm{ch}$ minus the number of parameters in all the model components combined.
For large $N_\mathrm{dof}$, the $\chi^2$ distribution approaches a Gaussian distribution with mean $N_\mathrm{dof}$ and variance $2N_\mathrm{dof}$. 
Therefore the reduced $\chi^2$ is approximately $\sim \mathcal{N}\left(1, \sqrt{2/N_\mathrm{ch}}\right)$.
We use this probability density function to define a cut-off in the reduced $\chi^2$ above which the difference between the data and the model fit becomes very large, and unlikely to be due to noise: this indicates that the model is a poor fit to the data.
The cut-off in the reduced $\chi^2$ that we chose includes all fits within the $5\sigma$ limits of the Gaussian distribution, and all fits with a reduced $\chi^2 < 1$.

Cross-validation provides an additional way for quantifying which model describes the data best (Colin Gillespie, private communication), but we will not consider this technique here.
\cite{raftery1995} proposed what he called Occam's window to reduce the computational load when averaging over a large ensemble of models. 
Since we only test small numbers of models, we do not apply Occam's window.

\section{Case study: Observations with the ATCA in the 16 cm band}\label{spass.sec}
To understand the strengths and weaknesses of automated fitting and selection of models, we run two series of tests with $\textsc{Firestarter}$, see Sections~\ref{MC_test_series1.sec} and \ref{MC_test_series2.sec}.
Each series investigates different aspects of the code: how often it identifies the injected model as the correct model to describe the data, how much more weight it gives to the preferred model compared to the other models being tested (the selectivity of the test), and how the recovered parameters relate to the injected parameters.
We will compare results for the AIC and BIC, with and without model averaging.
In each series of tests we simulate observations with the Australia Telescope Compact Array (ATCA) in the 16 cm band, covering a continuous frequency range between 1300-3124 MHz with 8-MHz channels. 
For each model we generate 1000 mock observations, adding Gaussian noise with a standard deviation $\sigma=1$ flux density units independently to Stokes $Q$ and $U$ in each channel.

\begin{table}
\centering
\caption{
Characteristics of the models used in Section~\ref{MC_test_series1.sec}. 
The first digit in the name of each test indicates the type of model from Section~\ref{continuous_model.sec} that was injected. 
$L_\mathrm{ref}$ refers to the value that we used for $|\bmath{L}_\mathrm{ref}|$ in equations~(\ref{point_source.eqn})-(\ref{polvectwo.eqn}). 
In all cases the emission has a spectral index $\alpha=0$, an intrinsic polarization angle $\chi_0 = 0\degr$, and RM$_0$ = 0 \radm. 
RM$_\mathrm{Rayleigh}$ = 71.4~\radm\ for the frequency setup that we simulated.
}
\label{overview_test1.tab}
\begin{tabular}{lcrcl}
\hline
Test & $L_\mathrm{ref}$ & \multicolumn{3}{c}{RM width} \\
 & (arbitrary units) & \multicolumn{3}{c}{(rad m$^{-2}$)} \\
\hline
\multicolumn{5}{l}{\emph{Single point source}}\\
101 & 10 & & 0 & \\
\\
\multicolumn{5}{l}{\emph{Models with a Gaussian RM distribution}}\\
201 & 50 & $\sigma_\mathrm{RM}$ & = & RM$_\mathrm{Rayleigh}$ \\
202 & 50 & $\sigma_\mathrm{RM}$ & = & RM$_\mathrm{Rayleigh}$/4 \\
203 & 50 &  $\sigma_\mathrm{RM}$ & = & RM$_\mathrm{Rayleigh}$/25 \\
\\
\multicolumn{5}{l}{\emph{Models with a rectangular RM distribution}}\\
301 & 25 & $\Delta\mathrm{RM}$ & = & 2 RM$_\mathrm{Rayleigh}$ \\
302 & 25 & $\Delta\mathrm{RM}$ & = & RM$_\mathrm{Rayleigh}$ \\
303 & 25 & $\Delta\mathrm{RM}$ & = & RM$_\mathrm{Rayleigh}$/4 \\
304 & 25 & $\Delta\mathrm{RM}$ & = & RM$_\mathrm{Rayleigh}$/25 \\
\hline
\end{tabular}
\end{table}

\begin{figure}
\begin{centering}
\resizebox{0.9\hsize}{!}{\includegraphics{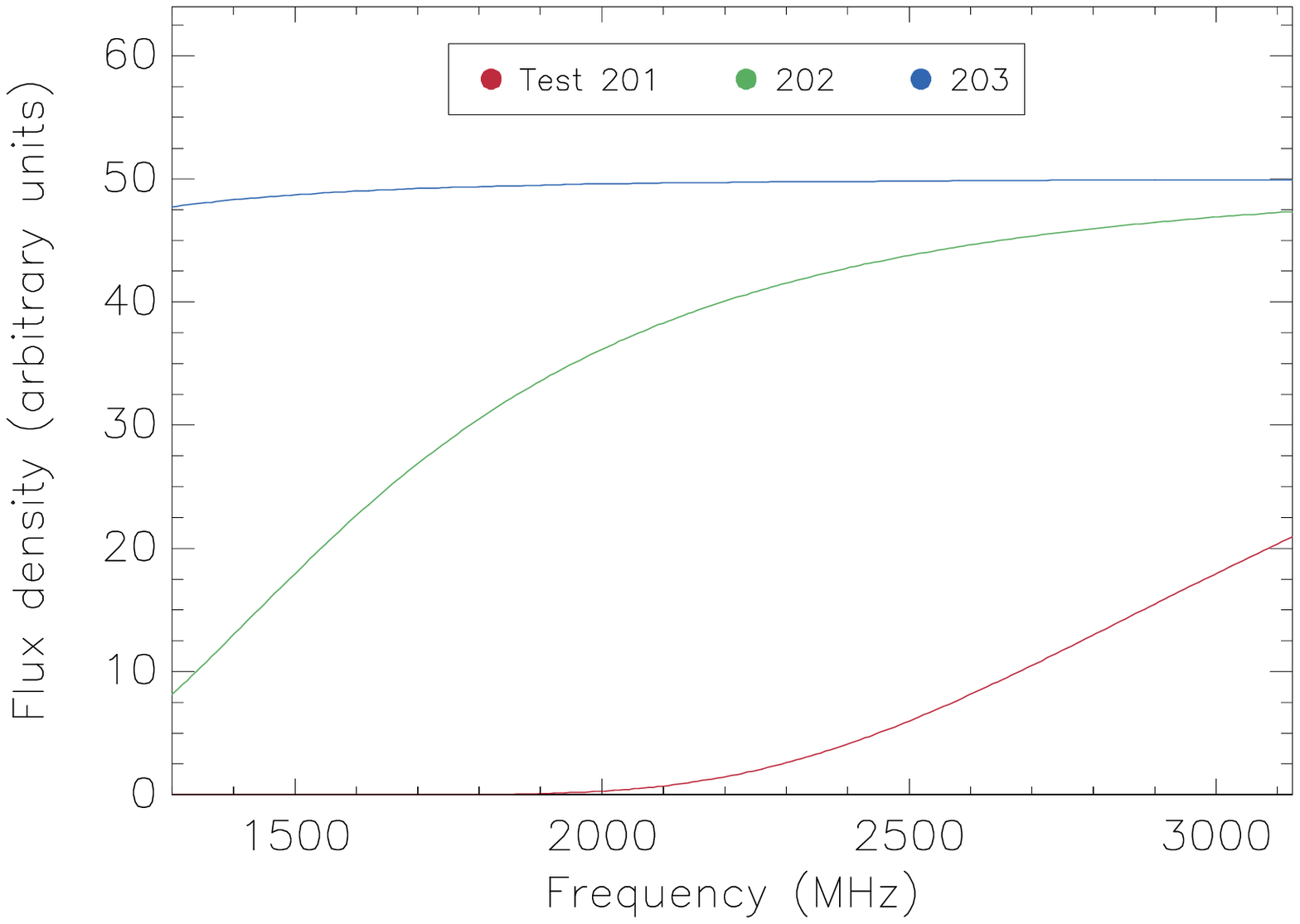}}
\resizebox{0.9\hsize}{!}{\includegraphics{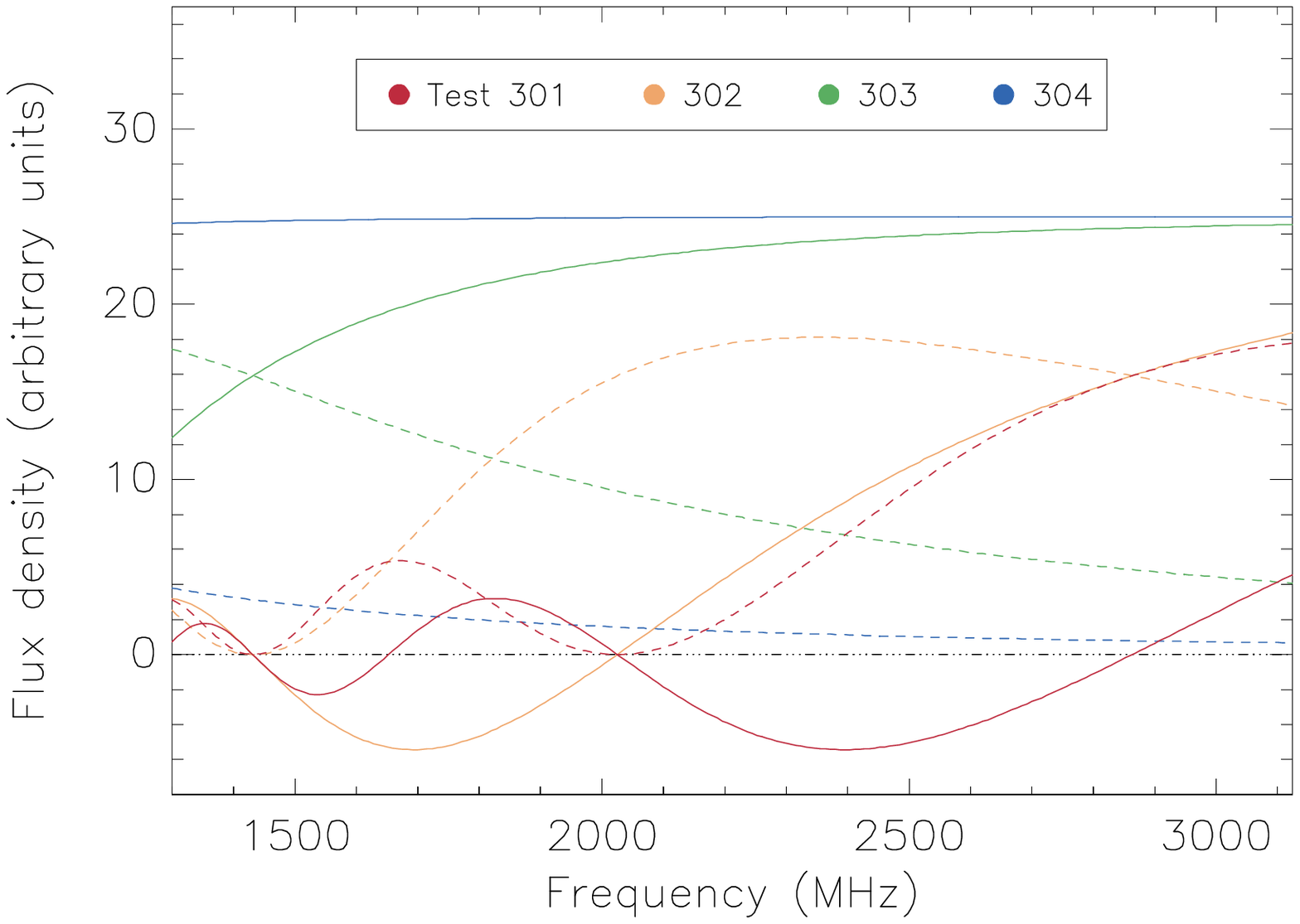}}
\caption{
Noise-free simulated spectra for tests where models~2 and 3 were used (top and bottom panels, respectively). Different models use different colours; Stokes $Q$ is shown with solid lines, Stokes $U$ with dashed lines. 
Table~\ref{overview_test1.tab} provides detailed information on the models we used.
The tests with Gaussian RM distributions do not have any emission in Stokes $U$, only noise, therefore we do not show Stokes $U$ profiles in the top panel.
}
\label{test_fspectra.fig}
\end{centering}
\end{figure}

Without prior knowledge, ranking models using the AIC or BIC requires fitting each of the models shown in Fig.~\ref{modelgrid.fig}; it is not sufficient to test only the more complex models.
Consider the situation where the source is described by a simple model that is \emph{not} in the list of models being tested (the `parent'), but we do test a model that has one additional parameter (the `daughter').
Furthermore, assume that the fitting procedure finds the exact parameters that describe the source, and a value of zero for the extra parameter that is in the daughter model but not in the parent model.
If we calculate the difference in AIC or BIC between the model that we do test, and the parent model that we do not test, we find a difference between the AIC values $\Delta_j = 2$, and for the BIC $\Delta_j \approx 6$ for the frequency setup we simulate.
These values for $\Delta_j$ are large enough that, if we had included the parent model in our test, then with one stroke it would have been ranked above its daughter model with a ``postive" to ``strong" preference (table~6 in \citealt{raftery1995}) based on its BIC, and a ``weak" to ``positive" preference based on its AIC.
Not including the simpler parent model in the ranking procedure therefore could lead to the wrong physical model being selected to describe the data.
This example demonstrates that the true model for the source has to be in the list of models being ranked. Therefore, when compiling a list of models that are to be tested, reliable, physically-motivated models should be used instead of toy models.

In the example from the previous paragraph, the parameters that are in common between the parent and daughter models have the same values, and the additional parameter in the daughter model is zero. 
This means that the reduced chi squared values for the fits are the same for parent and daughter: in this situation the reduced chi squared cannot be used to rank models. 
However, the reduced chi squared does provide an absolute measure for the quality of the fit, whereas the AIC and BIC only provide relative measures.

\subsection{Test series 1}\label{MC_test_series1.sec}
In the first series of tests we inject a single point source in RM, Gaussian RM distribution, or rectangular RM distribution (model types 1-3 from Section~\ref{continuous_model.sec}). 
Table~\ref{overview_test1.tab} lists the parameters that describe these models, and noise-free frequency spectra for these models are shown in Fig.~\ref{test_fspectra.fig}.
Different tests use different widths for the Gaussian and rectangular RM distributions, and/or different signal-to-noise ratios. 
For each test listed in Table~\ref{overview_test1.tab} we run two additional tests, in which we reduce the amplitude of the injected signal by a factor of ten and one hundred.
We include one test where no signal was injected at all.
The models we fit to the simulations consist of one or two components, each component could be any one of the eight models described in Section~\ref{continuous_model.sec}.
To find the starting point for the LM algorithm that is used by $\textsc{Firestarter}$, we calculated RM spectra out to $\pm$ 2500~\radm.

\begin{figure*}
\begin{centering}
\resizebox{0.145\hsize}{!}{\includegraphics{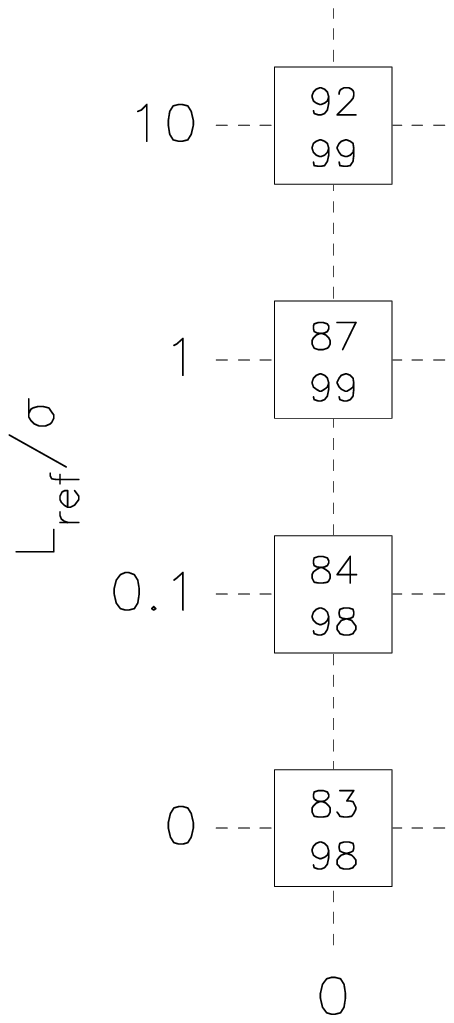}}
\resizebox{0.33\hsize}{!}{\includegraphics{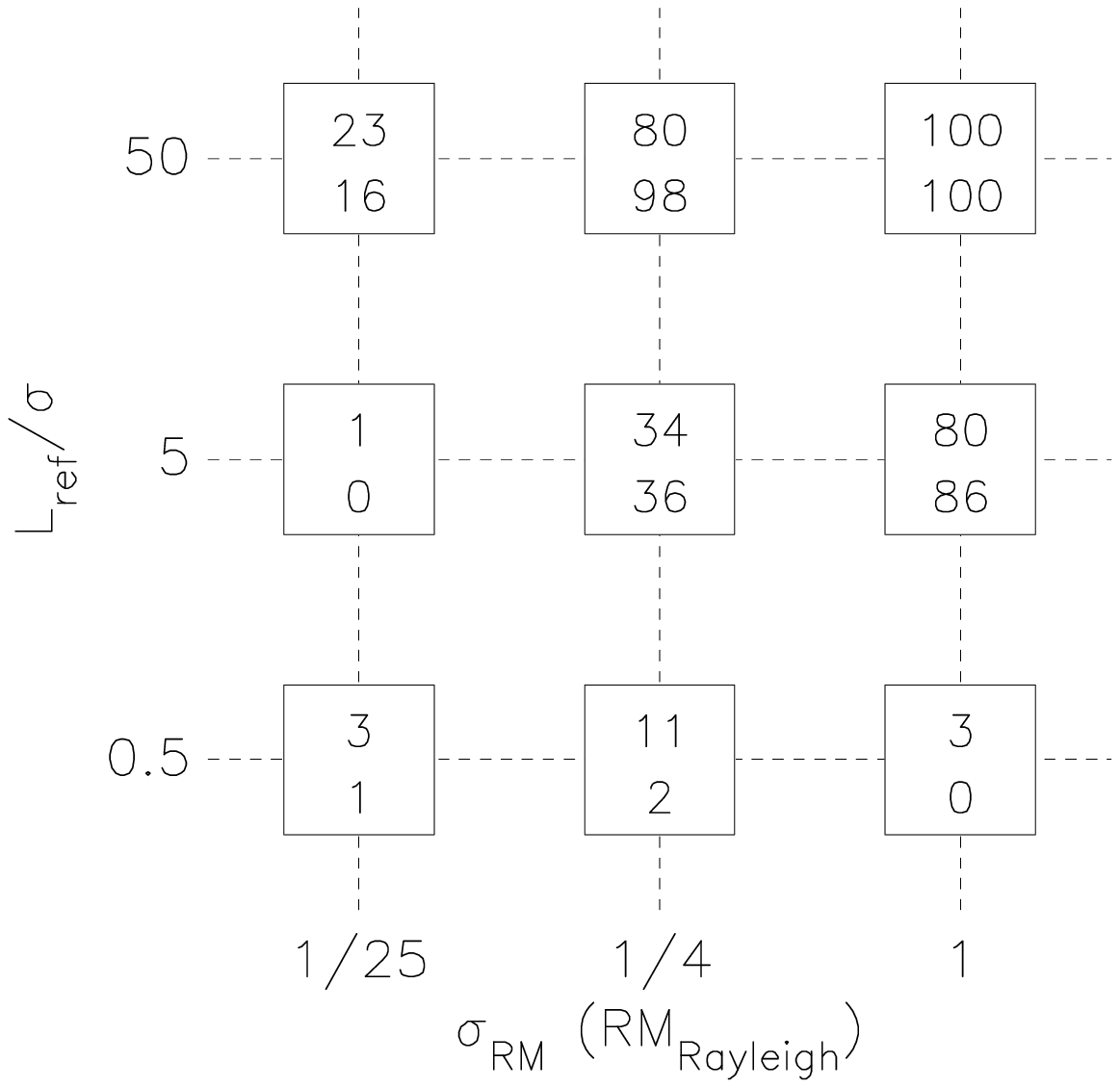}}
\resizebox{0.42\hsize}{!}{\includegraphics{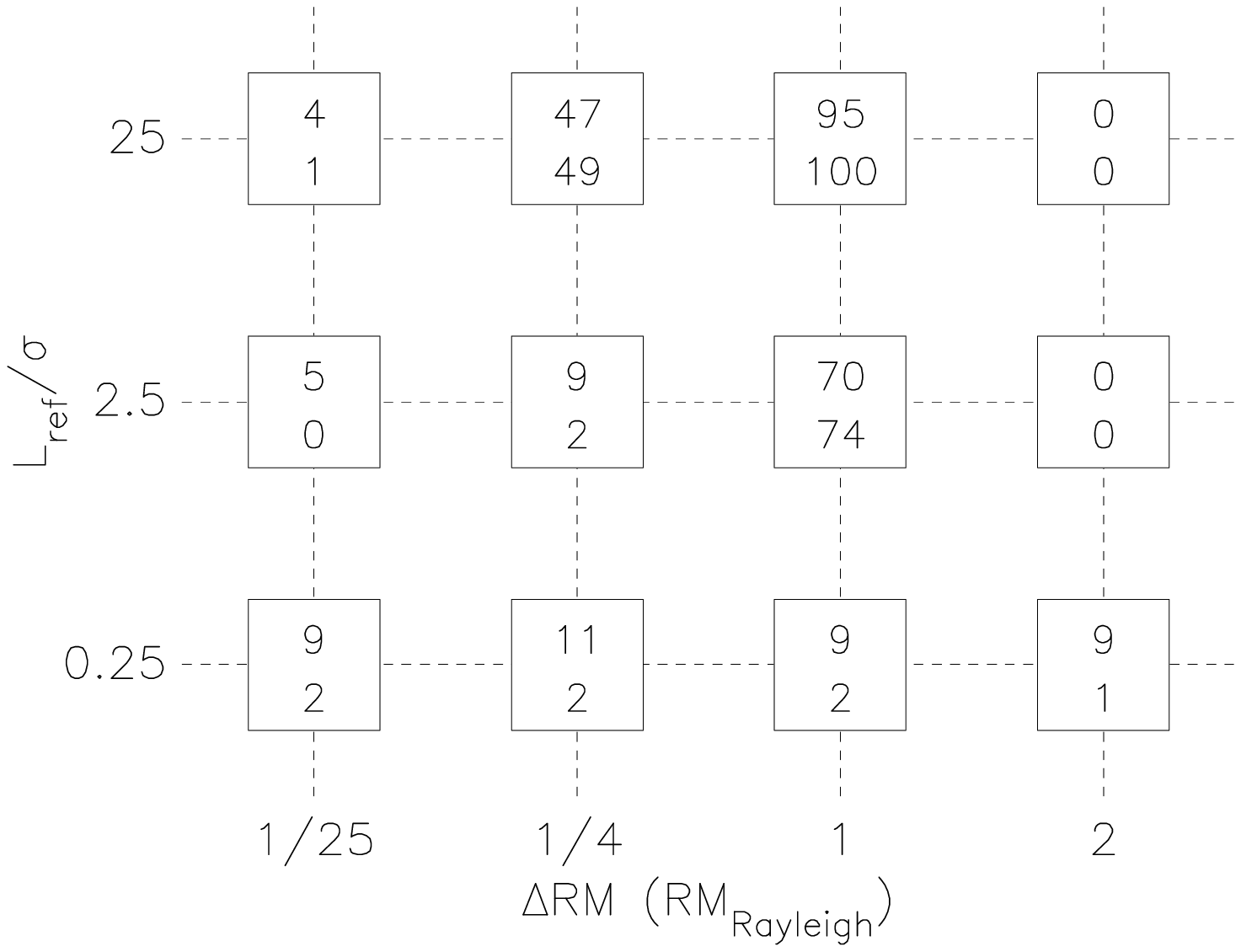}}
\caption{
Overview of the number of times, converted to percentages, the AIC and BIC identified the correct model, for different source models, different signal-to-noise ratios, and different widths of the RM distribution. 
The three panels show results for models of a point source (left), Gaussian RM distribution (middle), and rectangular RM distribution (right). 
For each grid point, the number at the top (/bottom) shows the percentage for the AIC (/BIC).
In the panel on the left, the numbers in the square at the bottom show how often (expressed as a percentage) the model for a point source was selected when there was no signal but only noise.
The results shown in this figure are for tests where we considered only models consisting of one component.
}
\label{aic_bic_idents_series1.fig}
\end{centering}
\end{figure*}

The Monte Carlo simulations show that the BIC identifies the correct model much more frequently than the AIC: Fig.~\ref{aic_bic_idents_series1.fig} shows the number of correct identifications by the AIC and BIC for the tests listed in Table~\ref{overview_test1.tab}, for different signal-to-noise ratios. 
Clearly, neither the AIC nor the BIC identifies the correct model all the time.
Such misidentifications lead to the source being interpreted with the wrong physical model.
When we allow for not one but also two model components then the number of correct identifications changes only slightly for the BIC (up to 5/1000 simulations) but reduces dramatically for the AIC (up to about 900/1000 simulations).
In all but one of the tests, the AIC typically prefers models with an additional point source component, which explains the many misidentifications if we fit models with up to two components. The BIC does not suffer from this effect.
If the signal is weak, sources are almost always misidentified as point sources.
Sources are also often misidentified if the RM range of the emission is very narrow, typically as point sources, whereas if sources emit over a wide range in RM they are strongly depolarized.
Therefore, sources can be identified correctly only if they emit over a range in RM that is neither too narrow, nor too wide.
Non-zero spectral indices will change the boundaries of this `Goldilocks region' in parameter space (see also \citealt{arshakian2011}).

\begin{figure}
\begin{centering}
\resizebox{0.9\hsize}{!}{\includegraphics{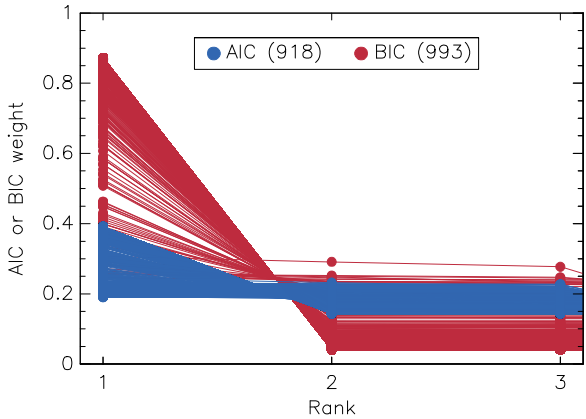}}
\caption{
Model weights for the top three models as ranked by the AIC (blue points) and BIC (red points). The panel at the top shows the number of times the AIC and BIC identified the correct model.
This figure shows results for test 101, where we simulated a point source with a signal-to-noise ratio $L_\mathrm{ref}/\sigma = 10$, and we fitted only models consisting of one source component.
}
\label{selectivity.fig}
\end{centering}
\end{figure}

The BIC not only identifies the correct model more frequently than the AIC, in our Monte Carlo simulations it typically also assigns a higher weight to the preferred model than the AIC (Fig.~\ref{selectivity.fig}).
This implies that the model that is preferred by the BIC has a higher probability of being the same as the injected model, compared to the model that is preferred by the AIC \citep{wassermann2000}: the BIC is more selective than the AIC.
This agrees with the analysis of Fig.~\ref{aic_bic_idents_series1.fig} that we presented in the previous paragraph.
Furthermore, the AIC weights for the top-three models are typically higher if we only fit models consisting of one component to the data than if we allow for up to two components; for the BIC weights this difference is less pronounced.
Such changes in the weights imply that models consisting of two components are assigned non-zero weights by the AIC, reducing the weights of the top-three models, and confirming that the AIC is less selective than the BIC.

\begin{figure}
\begin{centering}
\resizebox{\hsize}{!}{\includegraphics{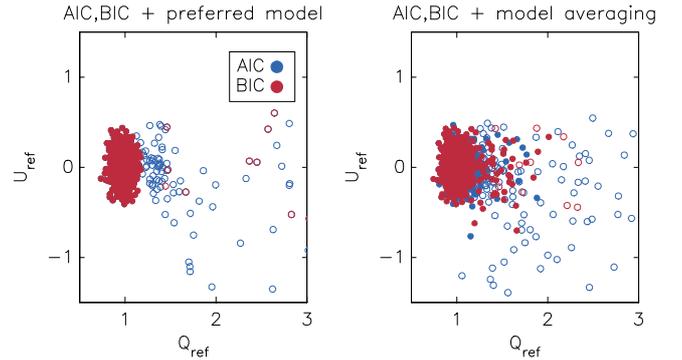}}
\caption{
Distribution of the values of $Q_\mathrm{ref}$ and $U_\mathrm{ref}$ that maximize the likelihood, considering only the model that is ranked highest by the AIC and BIC (left panel) or model averaging (right panel). 
In this test we simulated a single point source with $L_\mathrm{ref}/\sigma$=1, and we considered only source models that consist of a single component. 
Filled circles indicate simulations where the preferred model is the same as the injected model, open circles are misidentifications.
}
\label{AIC_BIC_Argand_diagrams.fig}
\end{centering}
\end{figure}

Based on the analysis we presented in SL17, for a point source that is observed at a high signal-to-noise level we expected to find Gaussian distributions for RM, $\alpha$, $Q_\mathrm{ref}$ and $U_\mathrm{ref}$, that are centred on the parameter values of the model we simulated.
Instead, we found that some simulations showed asymmetric parameter distributions, and non-zero values for the source parameters $\sigma_\mathrm{RM}$ and $\Delta\mathrm{RM}$, which should all be zero for a point source.
Fig.~\ref{AIC_BIC_Argand_diagrams.fig} illustrates that parameter distributions can become asymmetric if the model that is ranked highest by the AIC or BIC is not the same as the model we simulated.
Model averaging and misidentifications also explain why in some cases the maximum likelihood estimators for $\sigma_\mathrm{RM}$ and $\Delta\mathrm{RM}$ are not zero if we simulated a point source.
Our simulations of rectangular or Gaussian RM distributions are affected in similar ways.
By using the BIC without model averaging, and observing at a high signal-to-noise level, these adverse effects can be mitigated (but not nullified altogether).

In certain tests, $\textsc{Firestarter}$ systematically selects a model that is more complex than the model we injected. For example, in test 301 the BIC prefers model five instead of the simpler model three in about 750/1000 Monte Carlo simulations. Such behaviour could indicate that the program converged on a local maximum in the likelihood when it fitted the simpler model. 
This can be investigated by trying different starting points for the LM algorithm, and comparing the results: if the results agree, then the algorithm has (probably) found the global maximum in the likelihood. 
For the brightest point source model that we tested we changed the starting point of the LM algorithm;
this led to only minor differences in the results, model-averaged parameters being affected the strongest.

\begin{figure*}
\begin{centering}
\resizebox{0.8\hsize}{!}{\includegraphics{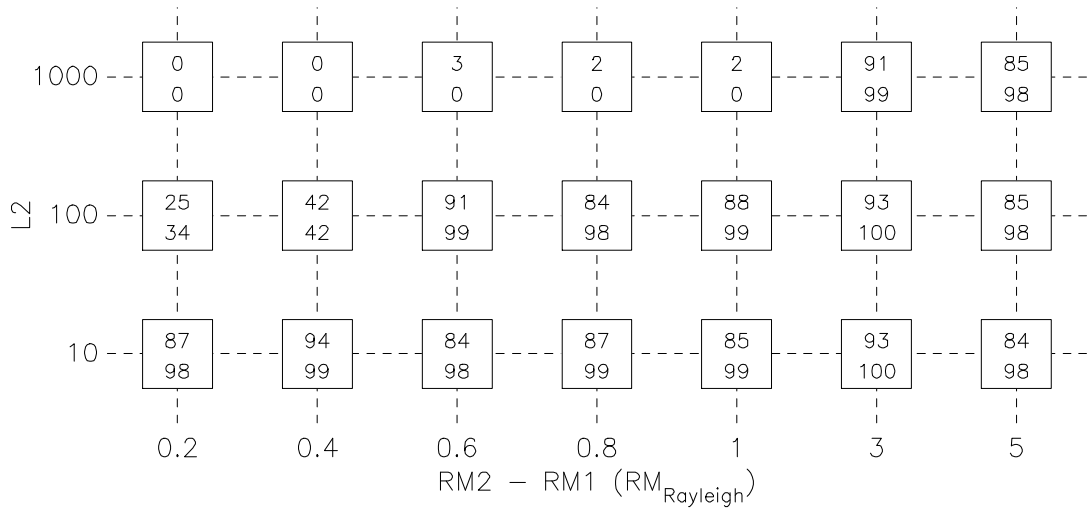}}
\caption{
Equivalent of Fig.~\ref{aic_bic_idents_series1.fig} for test series 2, where we consider models consisting of up to two components. For this particular set of simulations, the second point source emits at an intrinsic polarization angle $\chi_{0,2}=90\degr$.
}
\label{aic_bic_idents_series2.fig}
\end{centering}
\end{figure*}

\subsection{Test series 2}\label{MC_test_series2.sec}
In this series of tests we simulate two point sources, each of them bright enough so that the effects of noise bias are negligible. The first point source emits a polarized flux density of 1000 units (the standard deviation of the noise $\sigma=1$ flux density unit in each channel) at RM = 0~\radm, and has an intrinsic polarization angle $\chi_{0,1}=0\degr$. We vary the polarized flux density, RM, and intrinsic position angle of the emission from the second source: we consider polarized flux densities of 1000, 100, and 10 units, RMs between 0.2 and 5 times RM$_\mathrm{Rayleigh}$ (see Fig.~\ref{aic_bic_idents_series2.fig} for the exact values), and intrinsic polarization angles $\chi_{0,2}$ = 0\degr, 45\degr, 90\degr, and 135\degr. 
We fit models containing up to two components, each component is selected from the list of eight models that we presented in Section~\ref{continuous_model.sec}.

Fig.~\ref{aic_bic_idents_series2.fig} shows the number of simulations where the AIC or BIC identified the model we simulated correctly; in this case, the second source emits at $\chi_{0,2}=90\degr$.
What came as a surprise is that even bright, well-separated sources can be misinterpreted as other models (top row of squares in Fig.~\ref{aic_bic_idents_series2.fig}): in this case the model consisting of two point sources is typically misinterpreted as a rectangular $L(\mathrm{RM})$ distribution. Both models have very similar values for the reduced chi squared (they differ by about 0.01), and because the AIC and BIC penalize models with more parameters, a rectangular $L(\mathrm{RM})$ distribution is preferred over the model consisting of two point sources that we simulated. 
For the three other intrinsic polarization angles $\chi_{0,2}$ that we simulated, the BIC identifies the correct model in more than 96 per cent of the simulations if $\mathrm{RM}_2-\mathrm{RM}_1 > $ 0.6.

In the second row of boxes in Fig.~\ref{aic_bic_idents_series2.fig}, when $\mathrm{RM}_2-\mathrm{RM}_1$ equals 0.2 or 0.4, the AIC and BIC apparently identify the injected model correctly in fewer than half of our simulations. However, on closer inspection we found that in a substantial number of simulations the AIC and BIC preferred a more complex model that was fitted with $\Delta$RM or $\sigma_\mathrm{RM}$ equal to zero, which means that such a model would be interpreted physically as a point source. For the BIC this occurred in 340 simulations when $\mathrm{RM}_2-\mathrm{RM}_1$=0.2, and in 165 simulations when $\mathrm{RM}_2-\mathrm{RM}_1$=0.4 (166 and 354 of the simulations, respectively, if we consider the AIC instead of the BIC). 

\begin{figure}
\begin{centering}
\resizebox{0.8\hsize}{!}{\includegraphics{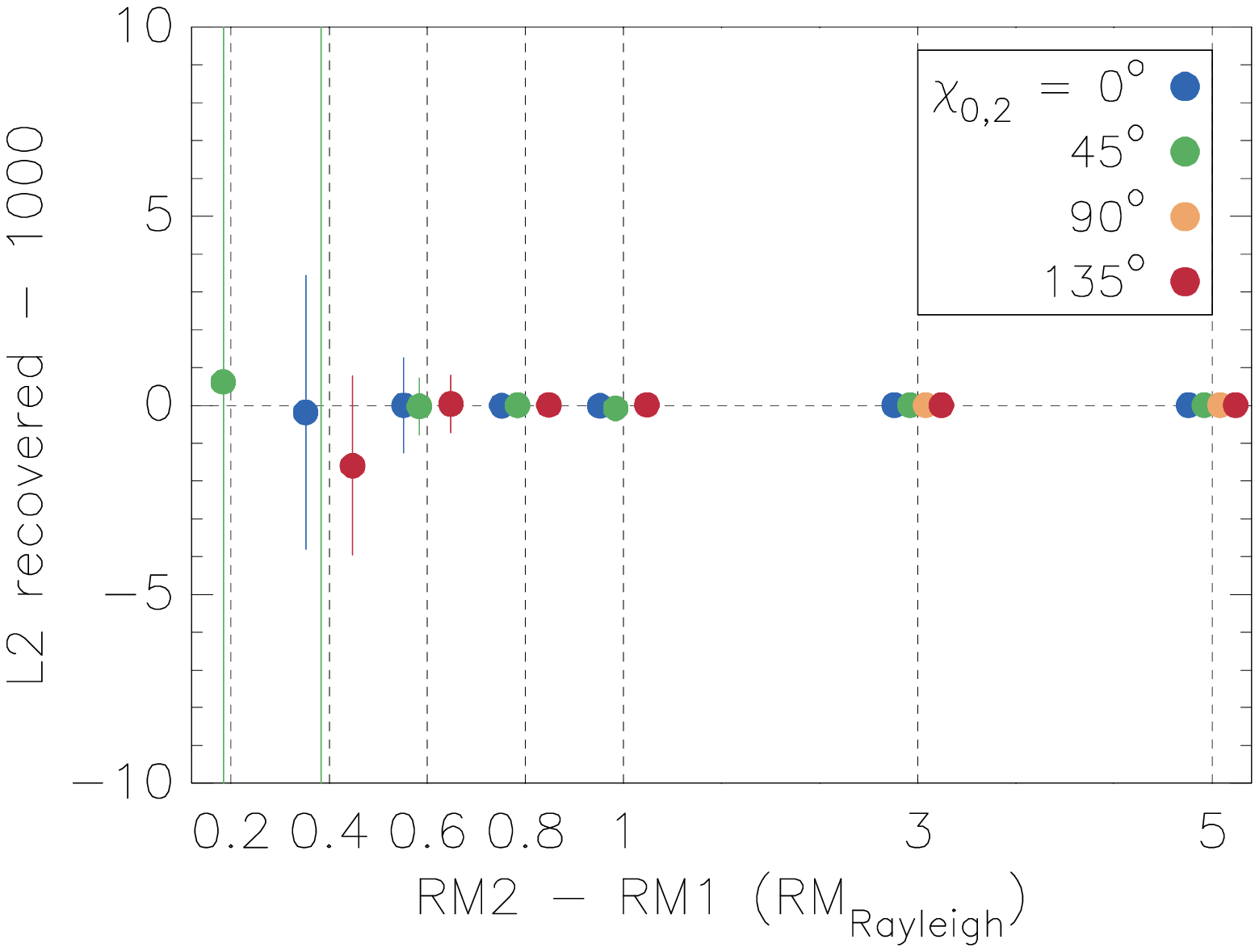}}
\resizebox{0.8\hsize}{!}{\includegraphics{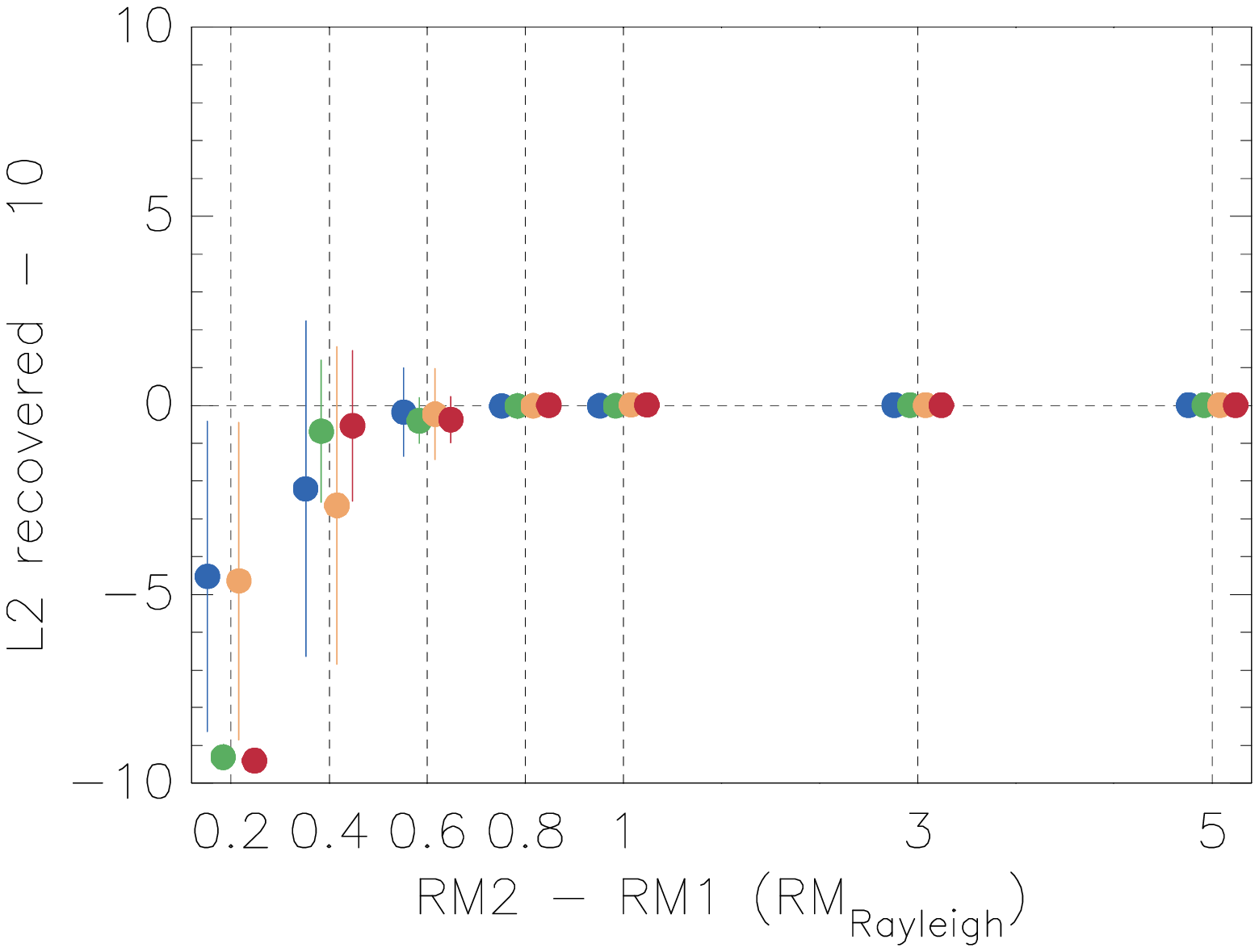}}
\caption{
Median values for the recovered polarized flux density of the second source component. 
The second source emits a polarized flux density of 1000 units (top panel) or 10 units (bottom panel), the difference between its RM and the RM of the first source at RM = 0~\radm\ is plotted along the x-axis. Different colours show results for different intrinsic polarization angles 
The error bars are calculated as 1.48 times the median absolute deviation, which is equal to the 1-sigma limit of a Gaussian distribution.
}
\label{Derived_L2.fig}
\end{centering}
\end{figure}

We analysed the source parameters found by $\textsc{Firestarter}$ also for the simulations in this test series. Fig.~\ref{Derived_L2.fig} shows the recovered polarized flux density for the second source component, other parameters behave in a very similar way. If the two source components are well separated, $|\mathrm{RM}_2 - \mathrm{RM}_1| \gg \mathrm{RM}_\mathrm{Rayleigh}$, then the polarized flux densities and RMs of the sources can be recovered without a problem. Only if the sources are separated in RM by less than about 0.6 $\mathrm{RM}_\mathrm{Rayleigh}$ do the source parameters deviate strongly from the injected values.
A complicating factor is that the algorithm does not always identify a two-component model as the best model to describe the data, as was already shown by the top row of panels in Fig.~\ref{aic_bic_idents_series1.fig}. This explains the absence of orange points in the top panel.

Previously, \cite{farnsworth2011} established that RM synthesis and RMClean sometimes misidentify double point sources in RM as a single point source. 
\cite{kumazaki2014} and \cite{miyashita2016} confirmed this in their analyses. 
Furthermore, Kumazaki et al. found that such misidentifications occur less frequently if the two point sources in the injected model have very different flux densities. 
The analysis we presented here and in Section~\ref{MC_test_series1.sec} confirms that also QU fitting is not immune to these effects.

Also in this series of tests $\textsc{Firestarter}$ sometimes prefers models that are more complex than the model we injected, hinting at the program converging on a local instead of the global maximum in the likelihood.

\section{Related topics}\label{related_topics.sec}
\subsection{Apodization}\label{apodization.sec}
To find a starting point for the LM algorithm, $\textsc{Firestarter}$ calculates an RM spectrum. 
This happens each time a new component is added to the model, therefore, fitting models consisting of many components is computationally expensive.
If the wings of the RMSF are suppressed, the RM spectrum would have to be recalculated only for a narrow range of RM values.
This is known as apodization in optical astronomy, and it can be accomplished by decreasing the weights of channels at the edges of the frequency band.

\begin{figure}
\resizebox{\hsize}{!}{\includegraphics{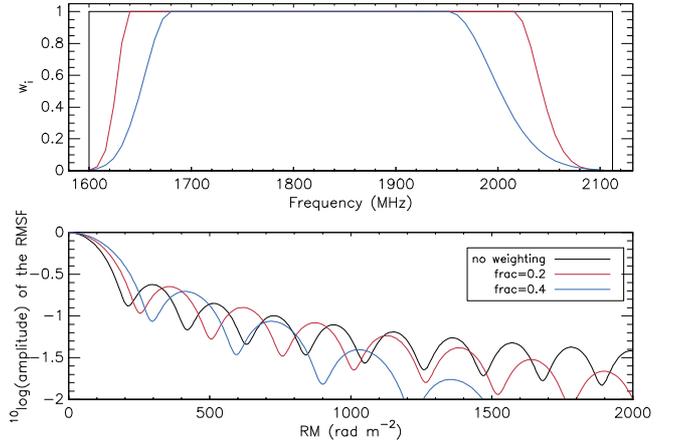}}
\caption{RMSFs with and without applying a Gaussian window function in wavelength squared. The top panel shows the different weighting functions as a function of frequency, while the bottom panel shows the weighted RMSF for each window function. Red and blue lines correspond to sacrificing 20\% and 40\%, respectively, of the wavelength squared coverage of the simulated observations. We assume $\alpha=0$, and that the noise variances in Stokes $Q$ and $U$ are equal and constant across the frequency band.
}
\label{apodization.fig}
\end{figure}

Fig.~\ref{apodization.fig} illustrates apodization for three different strengths of the Gaussian window function; \cite{harris1978} discuss alternative choices.
To calculate these spectra, we divided the noise variances of the individual channels, $\sigma_{Q,i}^2$ and $\sigma_{U,i}^2$, by weights $w_i$ that we defined as follows.
Let $\lambda^2_\mathrm{min}$ and $\lambda^2_\mathrm{max}$ be the frequency range of the observations converted to units of wavelength squared. 
If the taper takes up a fraction `frac' of the frequency band, then the weights $w_i$ change smoothly between zero and one on either side of the band over a distance $\delta\lambda^2=\left(\mathrm{frac}/2\right)\times\left(\lambda^2_\mathrm{max}-\lambda^2_\mathrm{min}\right)$, and are described by
\begin{eqnarray}
w_i =
\begin{cases}
\exp\left(-0.5\left(\left[\lambda_i^2 - \left(\lambda^2_\mathrm{min}+\delta\lambda^2\right)\right]\big/ \sigma_{\lambda^2}\right)^2\right) \\
\exp\left(-0.5\left(\left[\lambda_i^2 - \left(\lambda^2_\mathrm{max}-\delta\lambda^2\right)\right]\big/ \sigma_{\lambda^2}\right)^2\right)\, ,
\end{cases}
\label{gaussian_window.eqn}
\end{eqnarray}
for the high-frequency and low-frequency edges of the band, respectively, where $3\sigma_{\lambda^2}=\delta\lambda^2$ (top panel of Fig.~\ref{apodization.fig}).

Apodization affects RM synthesis and QU fitting in the same way, because the point spread function is described by the same equation in both cases (for RM synthesis this is equation~37 from \citealt{brentjens2005}, and for QU fitting this is the part from equation~(\ref{beat.eqn}) which describes the contribution from source two).
However, apodization does not make optimal use of the information available in the frequency channels for which the weights are reduced.
This increases both the noise level (and therefore the measurement uncertainties) and the width of the main peak in the point spread function, as Fig.~\ref{apodization.fig} shows.

Future applications of apodization could include changing the weights $w_i$ to reflect the local density of data points per unit wavelength squared, to introduce the concept of robust weighting as used in radio interferometry \citep{briggs1995} to QU fitting and RM synthesis.

\subsection{Observing time requirements in QU fitting and RM synthesis}\label{sensitivity.sec}
If RM synthesis produces source parameters that also maximize the likelihood (see Section~\ref{rayleigh.sec}), then, for this set of parameters, both RM synthesis and QU fitting give the same log likelihood ratio, and therefore, the same detection significance (see section~2.4 in SL17).
Put in another way: under these conditions RM synthesis and QU fitting benefit in the same way from combining many frequency channels, potentially having a low signal-to-noise ratio, to detect a signal.
In both cases it is not necessary to integrate for long periods of time to detect a source in individual frequency channels.
However, it becomes easier to identify the correct source model in an automated ranking procedure if the signal is strong. If the signal is weak, then several physical models can describe the data equally well.
Also, we showed in \cite{schnitzeler2015b} that different models can produce frequency spectra that differ only subtly. These models can be told apart only if the source is detected in individual frequency channels with a high signal-to-noise ratio.
Programs like $\textsc{Firestarter}$ can be used to simulate ensembles of noise realisations for different source models, to determine the observing time required for identifying the correct model in (for example) 95\% of the realisations.
This can potentially lead to large savings on time requests in observing proposals.

\subsection{RMClean: when to stop adding components, and whether the cleaned RM spectrum is unique}\label{rmcleanstop.sec}
Typically RMClean is stopped when the polarized flux density of the highest peak in the residual RM spectrum is a few times the noise level: this avoids cleaning peaks in the RM spectrum that are produced by noise, saving computing resources.
$\textsc{Firestarter}$ fits a new source component only if the highest peak in the residual RM spectrum has a signal-to-noise ratio $>$ 1.5; we explain how we calculate this ratio in Appendix~\ref{signal-to-noise.sec}.
The calculation of the signal-to-noise ratio simplifies if the noise variances in Stokes $Q$ and $U$ are equal in each channel, $\sigma_{Q,i}^2=\sigma_{U,i}^2 \equiv \sigma_{L,i}^2$. 
Then the variances in the ML estimators of Stokes $Q$ and $U$ at the reference frequency, $\mathrm{var}\left(\hat{Q}_\mathrm{ref}\right)$ and $\mathrm{var}\left(\hat{U}_\mathrm{ref}\right)$, are given by:
\begin{eqnarray}
\mathrm{var}\left(\hat{Q}_\mathrm{ref}\right) = \mathrm{var}\left(\hat{U}_\mathrm{ref}\right) = 1\bigg/\sum_{i=1}^{N_\mathrm{ch}} \frac{1}{\sigma_{L,i}^2}\left(\frac{\nu_i}{\nu_\mathrm{ref}}\right)^{2\hat{\alpha}}\, , 
\end{eqnarray}
and cov($\hat{Q}_\mathrm{ref},\hat{U}_\mathrm{ref}$) = cov($\hat{U}_\mathrm{ref},\hat{Q}_\mathrm{ref}$) = 0 (Note that in SL17 these equations are missing the factor that contains the ML estimator of $\alpha$: see the erratum for that paper). 
Since we are looking for the highest peak in the residual RM spectrum, which means confining our search to $\alpha=0$, $\hat{\alpha}=0$ in the expressions for $\mathrm{var}\left(\hat{Q}_\mathrm{ref}\right)$ and $\mathrm{var}\left(\hat{U}_\mathrm{ref}\right)$. 

Under the conditions outlined in section~2.3 in SL17, these are also the variances and covariances in Stokes $Q$ and $U$ at the RM that produces the highest peak in the RM spectrum.
If $L$ would follow a Rayleigh distribution, then only 33\% of noise realisations would have a signal-to-noise ratio larger than 1.5.
However, the $L$ of the highest peak in an RM spectrum does not follow a Rayleigh distribution, even if this peak is produced by noise, as \cite{hales2012} and \cite{macquart2012} showed. Only if we consider the distribution of $L$ produced by noise at a single RM, does $L$ follow a Rayleigh distribution.
Furthermore, equation~17 from SL17 was derived for emission by a single point source and not for sources with multiple components (see also Section~\ref{rayleigh.sec}).
Nevertheless, we choose this cutoff in polarized flux density for $\textsc{Firestarter}$ for practical reasons, even if this means that also peaks produced by noise will be fitted.
If adding additional components improves the quality of the fit only slightly (for example, because the additional components have very small polarized flux densities), then these models will be penalized by the AIC and BIC for having too many parameters.

If the noise variances in Stokes $Q$ and $U$ are not equal in each frequency channel, then there is no unique way in which the RM spectrum can be cleaned. 
By working out the variances and covariances in equation~(17) from SL17 one can show that in this case confidence regions in the $Q,U$ plane are no longer circles but ellipses (the Rayleigh distribution is replaced by the Hoyt distribution, \citealt{hoyt1947}) at the RM that produces the highest peak in the RM spectrum.
Furthermore, these confidence regions do not have to align with the $Q,U$ coordinate axes.
Therefore, when cleaning two peaks with similar RMs, the order in which the peaks are cleaned (start with the peak with the highest $L$ or the highest $L/\sigma$?) influences the outcome of RMClean: since the peaks have similar RMs, cleaning one will affect the other.
If the noise variances in $Q$ and $U$ in each channel are equal, then the signal with the highest $L$ is automatically also the signal with the highest $L/\sigma$, and the order in which to clean peaks in the RM spectrum is determined uniquely.

\subsection{Further considerations}\label{considerations.sec}
It is possible that several models have very similar values for the AIC and BIC, which means that the data can be explained using different physical models. For example, it is difficult to tell a Gaussian $L(\mathrm{RM})$ distribution from a point source if $\sigma_\mathrm{RM,external}$ is small: in that case one cannot conclude from the data if there is a turbulent depolarizing screen in the foreground.
Re-observing such a source can make a better identification possible because it increases the signal-to-noise ratio. Also, one could choose a different frequency setup that allows one to resolve narrow features in $L(\mathrm{RM})$.

Typically, astrophysical sources are very complex, and they might not be described by any of the simple models or combination of models from Section~\ref{continuous_model.sec}.
More advanced, physically-motivated models have been developed to explain the observations of a number of galaxies and AGN (see, for example, \citealt{laing2008} and \citealt{broderick2009}), and we strongly recommend that such models are included in future analyses based on QU fitting.

\section{Summary}\label{conclusions.sec}
We introduced an algorithm called $\textsc{Firestarter}$ that is based on QU fitting, and we identified issues with RM synthesis and RMClean that QU fitting does not suffer from: QU fitting can take into account the variation in sensitivity across the observing band, the spectral indices of multiple sources inside a single pixel, and it handles emission at comparable RMs better than RM synthesis. QU fitting-based algorithms like $\textsc{Firestarter}$ can decompose a measured signal as a series of point sources, similar to the combination of RM synthesis and RMClean, without suffering from the issues we identified in those two methods. 

Out of the methods for ranking models that we tested, the Bayesian Information Criterion without applying model averaging performs best. It identifies the correct (injected) model more frequently, and assigns a higher model weight to this model than its competitors. Misidentifications and model averaging often lead to asymmetric parameter distributions in cases where one would expect these to be symmetric. However, automated model fitting and selection is not perfect: we identified geometries where two bright point sources are misidentified in almost 100 per cent of the Monte Carlo simulations we ran.

We quantified whether polarized emission from a source is `resolved' in a way similar to the Rayleigh criterion in optics. An analysis of Monte Carlo simulations where we injected different source types shows that the injected model is identified correctly only if the range in RM across which the source emits is neither too wide (leading to strong depolarization) nor to narrow (when typically a point source model is preferred by $\textsc{Firestarter}$): we referred to this as the `Goldilocks region'. 

We showed that it is not sufficient to fit only the most complex model in QU fitting. Because metrics like the Bayesian Information Criterion penalize models with many parameters more severely, one should also fit simple models that can be created by setting at least one of the parameters of a complex model equal to zero. We encourage the use of more reliable physical models in QU fitting: models that are currently being used (can) oversimplify actual astrophysical sources.

Finally, we describe how RM spectra can be apodized to reduce sidelobes. We show that observing time requirements in QU fitting and RM synthesis are the same: in both methods the noise level is reduced by combining all the frequency channels across the band. Therefore, there is no need to detect a signal in individual frequency channels in QU fitting, as long as the signal can be detected reliably by combining all frequency channels. Programs like $\textsc{Firestarter}$ can be and perhaps should be used to determine the observing time required to detect a particular source type.
We discuss when $\textsc{Firestarter}$ and RMClean can be stopped, and show that if the noise levels in Stokes $Q$ and $U$ are not equal across the band, then there is no unique way for cleaning RM spectra.

\section*{Acknowledgements}
We would like to thank KJ Lee (Kavli Institute for Astronomy and Astrophysics at Peking University), Olaf Wucknitz and Aritra Basu (both at the Max Planck Institute for Radio Astronomy) and Andrew Fletcher (Newcastle University) for many fruitful discussions. 
We also thank the referee for their constructive comments that helped improve the manuscript.
We thank Glennys Farrar (New York University) for sharing her insights in fitting models with many parameters.
We also thank John Antoniadis from the Dunlap Institute for Astronomy \& Astrophysics (currently the Max Planck Institute for Radio Astronomy), Colin Gillespie from Newcastle University, Michele Vallisneri from NASA's Jet Propulsion Laboratory, and Golam Shaifullah from Bielefeld University (currently ASTRON in the Netherlands), for sharing their insights on the AIC and BIC. 
 
\bibliography{ne6e} 

\appendix
\section{Calculating the signal-to-noise ratio in the $Q,U$ plane}\label{signal-to-noise.sec}
\begin{figure}
\begin{centering}
\resizebox{0.6\hsize}{!}{\includegraphics{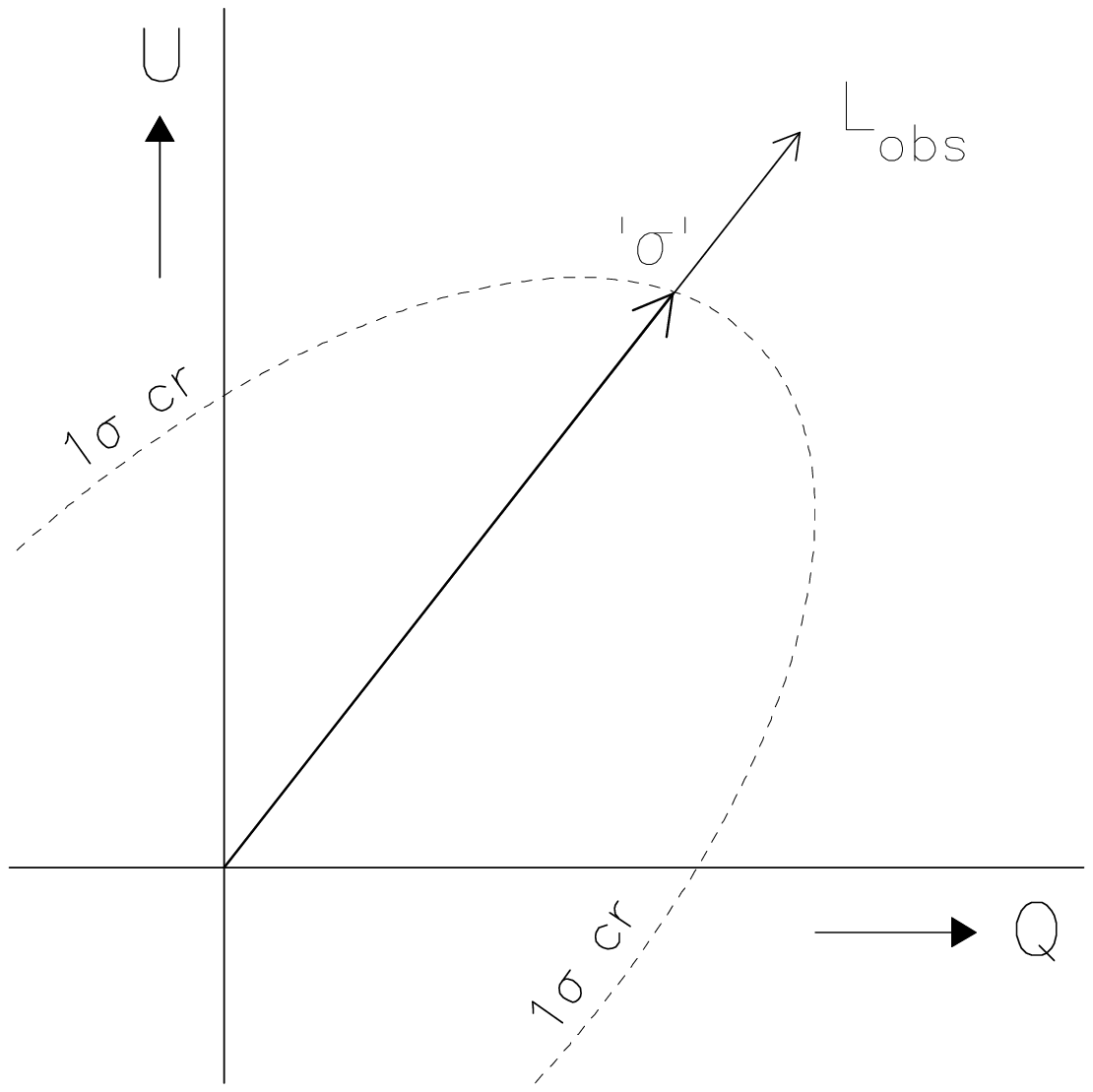}}
\caption{
Definition of the noise level associated with the observed polarization vector $\bmath{L}_\mathrm{obs}$.
In general, confidence regions in the ($Q,U$) plane are ellipses that are tilted with respect to the coordinate axes. 
In this example, the dashed line labeled `1$\sigma$ cr' shows the boundary of the 1-sigma confidence region.
To calculate the signal-to-noise level of the observation, we divide the length of the observed polarization vector $\bmath{L}_\mathrm{obs}$ by the extent of the 1$\sigma$ confidence region in the direction of $\bmath{L}_\mathrm{obs}$ (thick black vector, labeled `$\sigma$').
}
\label{snr_ellipse.fig}
\end{centering}
\end{figure}

To calculate the signal-to-noise ratio of a polarization vector $\bmath{L}_\mathrm{obs}$ we compare its length with the extent of the 1-sigma confidence region in the $Q,U$ plane. 
The covariance matrix from equation~(17) in SL17 describes the properties of these confidence regions if the noise is Gaussian.
If the covariance matrix can be written as a number times the identity matrix, then the confidence regions are circles, and in that case the signal-to-noise ratio is the length of the polarization vector divided by the radius of the circle.
However, in general the covariance matrix cannot be written this way, which means that confidence regions in the Stokes $Q,U$ plane are ellipses that do not align with the coordinate axes. 
Then the signal-to-noise ratio is equal to the length of the polarization vector divided by the extent of the 1-sigma confidence region in the direction of the polarization vector (Fig.~\ref{snr_ellipse.fig}).

To calculate this signal-to-noise ratio, we first derotate the confidence region so that it aligns with the $Q,U$ coordinate axes, then we apply a scaling so that the 1-sigma confidence region becomes the unit circle. This reduces the general situation to the simpler situation if the covariance matrix would have been the identity matrix, and the signal-to-noise ratio is then simply the length of the (derotated and rescaled) polarization vector.
The eigenvalues and eigenvectors of the covariance matrix make up the rotation and scaling matrices that we need for these transformations. The normalized eigenvectors are the columns in the rotation matrix $\mathbfss{R}$, and the scaling matrix $\mathbfss{S}$ is a diagonal matrix with the square roots of the eigenvalues alongs its diagonal (these properties are also used in whitening transformations and principal component analysis, see, e.g., \citealt{McDonough1995}). Then the signal-to-noise ratio is the length of the vector $\mathbfss{(R S)}^{-1}\bmath{L}_\mathrm{obs}$.

\bsp	
\label{lastpage}
\end{document}